%% file: paper.tex
\documentclass[useAMS,usenatbib]{mn2e}
\usepackage{graphicx}
\usepackage{setspace}
\usepackage{natbib}
\usepackage{color}
\usepackage{amsmath}
\usepackage{times}
\usepackage{aas_macros}

\title[Rotation of Halo Populations in the Milky Way and M31]{Rotation
  of Halo Populations in the Milky Way and M31}

\author[A. J. Deason, V. Belokurov, N. W. Evans]
{A.J. Deason$^{1}$\thanks{E-mail:ajd75,vasily,nwe@ast.cam.ac.uk},
V. Belokurov$^{1}$ and N. W. Evans$^{1}$ \\
$^{1}$Institute of Astronomy, Madingley Rd, Cambridge, CB3 0HA}

\begin{document}

\date{Accepted 2010 September 28. Received 2010 September 27; in original
  form 2010 May 05}

\pagerange{\pageref{firstpage}--\pageref{lastpage}} \pubyear{2010}

\maketitle

\label{firstpage}

\begin{abstract} 
  We search for signs of rotation in the subsystems of the Milky Way
  and M31 that are defined by their satellite galaxies, their globular
  cluster populations, and their BHB stars. A set of simple distribution functions are introduced to
  describe anisotropic and rotating stellar populations embedded in
  dark haloes of approximate Navarro-Frenk-White form. The BHB stars
  in the Milky Way halo exhibit a dichotomy between a prograde
  rotating, comparatively metal-rich component ($[\mathrm{Fe/H}] >
  -2$) and a retrograde rotating, comparatively metal-poor
  ($[\mathrm{Fe/H}] < -2$) component. The prograde metal-rich
  population may be associated with the accretion of a massive
  satellite ($\sim 10^9 M_\odot$). The metal-poor population may
  characterise the primordial stellar halo and the net retrograde
  rotation could then reflect an underestimate in our adopted local
  standard of rest circular velocity $\Theta_0$. If $\Theta_0$ is
  $\approx 240 \,\mathrm{kms^{-1}}$ then the metal-poor component has
  no rotation and there is a net prograde rotation signal of $\approx
  45 \, \mathrm{kms^{-1}}$ in the metal-rich component.  There is reasonable evidence that the
  Milky Way globular cluster and satellite galaxy systems are rotating
  with $\langle v_{\phi} \rangle \approx 50 \, \mathrm{kms^{-1}}$ and
  $\langle v_{\phi} \rangle \approx 40 \,\mathrm{kms^{-1}}$
  respectively. Furthermore, a stronger signal is found for the
  satellite galaxies when the angular momentum vector of the
  satellites is inclined with respect to the normal of the disc. The
  dwarf spheroidal satellites of M31 exhibit prograde rotation
  relative to the M31 disc with $\langle v_{\phi} \rangle \approx 40
  \,\mathrm{kms^{-1}}$. We postulate that this group of dwarf
  spheroidals may share a common origin. We also find strong evidence
  for systemic rotation in the globular clusters of M31 particularly for the
  most metal-rich.
\end{abstract}


\input{introduction}

\input{dfs}

\input{applications}

\input{conclusions}

\section*{Acknowledgements}
We thank the anonymous referee for a thorough reading of this
manuscript and for very useful comments. AJD thanks the Science and Technology Facilities Council (STFC) for
the award of a studentship, whilst VB acknowledges financial support
from the Royal Society.
\label{lastpage}

\bibliography{mybib}
\end{document}

%% file: introduction.tex
\section{introduction}

Our own Milky Way Galaxy and its nearest neighbour Andromeda (M31)
provide testing grounds for theories of galaxy formation and
evolution. Studies of local stellar populations explore the complex
processes which built up the galaxies.  The orbits of tracer
populations bear the imprint of the infall and subsequent accretion
history of the dark matter halo. In a monolithic collapse
model~\citep[see e.g.,][]{els}, the halo and the disc of the galaxy
are drawn from the same population. Any rotational signature in the
outer halo is thus aligned with the angular momentum of the
disc. However, in a hierarchical picture, substructure arriving late
at large radii may have little connection to the disc. Its evolution
is further complicated by processes such as tidal disruption, disk
shocking and dynamical friction. For example, prograde rotating
systems are more susceptible to dynamical friction than retrograde
(\citealt{quinn86}). \cite{norris89} suggested that the subsequent
disintegration of these accreted fragments by tidal forces may create
a net retrograde asymmetry in the remnant stellar halo population. The
presence, or absence, of a rotating signal can provide key insights
into the formation history of the halo.

There have been a number of previous kinematic studies of halo
populations in the Milky Way Galaxy. For example, \citet{frenk80}
found that the globular cluster system has a velocity distribution
that is isotropic about a net prograde rotation ($v_{\mathrm{rot}}
\sim 60 \: \mathrm{kms^{-1}}$). \cite{zinn85} extended this work by
restricting attention to halo globular clusters with $\mathrm{[Fe/H]}
< -1$ and found a slightly weaker net prograde motion
($v_{\mathrm{rot}} \sim 50 \: \mathrm{kms^{-1}}$). \cite{norris86}
used a sample of metal-poor objects ($\mathrm{[Fe/H]} < -1.2$) and
found evidence for weak prograde rotation, although he found no
kinematic difference between the clusters and halo stars, suggesting
these are part of the same population.

The isotropic velocity distribution of the globular clusters contrasts
with the highly anisotropic distributions usually found for Population
II stars (\citealt{woolley78}; \citealt{hartwick83}; \citealt{rat85}).
Cosmological simulations generally predict increasing radial
anisotropy with Galactocentric radii for halo stars
(e.g. \citealt{abadi06}), as a consequence of the accretion of
infalling 
structure. Observationally, the situation is less
clear-cut with radially biased (e.g. \citealt{chiba00}), isotropic
(e.g. \citealt{sirko04b}) and even tangentially biased
(\citealt{sommer-larsen97}) velocity distributions claimed for the
outer stellar halo.

More recently, a number of authors have advocated a dual halo
structure, which suggests that the stellar halo was formed by at least
two distinct phases of accretion events. The evidence includes studies
of the spatial profiles (\citealt{hartwick87}; \citealt{kinman94};
\citealt{miceli08}), as well as kinematic studies indicating net
retrograde rotation in the outer parts (\citealt{majewski92};
\citealt{carney96}, \citealt{wilhelm96}; \citealt{kinman07}). For
example, \cite{carollo07} claimed a dual halo structure consisting of
a weakly prograde rotating inner halo and a retrograde rotating outer
halo of lower metallicity ($\mathrm{[Fe/H]} < -2$). However, this idea
is inconsistent with work by \cite{chiba00} on local neighbourhood
stars and by \cite{sirko04b} on distant BHB stars who both find no evidence for
rotation in the outer halo.

This somewhat confused picture is partly caused by the difficulties of
the task. Full kinematic analyses of tracer populations are hampered
by small sample sizes and lack of full information on the phase space
coordinates. Often, work has been confined to comparatively local
samples, which may not be unbiased tracers of the overall
population. This, though, is beginning to change with more recent
studies on BHB stars reaching out to distances of $\sim 80$ kpc
(e.g. \citealt{sirko04b}; \citealt{xue08}), albeit with only the
line-of-sight velocity available. In fact, several surveys have the
promise to provide much larger samples of halo tracers with
well-defined kinematical parameters. This includes ongoing projects
like the Sloan Digital Sky Survey (SDSS, \citealt{york00}), and the
Sloan Extension for Galactic Understanding and Exploration (SEGUE,
\citealt{newberg03}), as well as future projects like the \emph{Gaia}
satellite (\citealt{turon05}), and the Large Synoptic Survey Telescope
(LSST, \citealt{ivezic08}).

There has also been a lively debate in recent years as to the
possibility that the satellite galaxies of the Milky Way lie in a
rotationally-supported disc. The idea may be traced back to
\citet{lynden-bell83}, who first suggested the Magellanic Clouds, Ursa
Minor and Draco may have been torn from a single progenitor, a
gigantic gas-rich proto-Magellanic Cloud. Working with more complete
datasets, \cite{kroupa05} claimed that satellite galaxies occupy a
highly inclined disc and that this is at odds with the predictions of
cosmological simulations. The available proper motion measurements
have been used to constrain the angular momenta orientations of the
satellites (e.g. \citealt{palma02}; \citealt{metz08}), indicating
evidence for some sort of coherent motion, possibly rotation. Even if
true, it remains disputatious as to whether this can be reconciled
within the cold dark matter framework of structure
formation~\citep{metz08,libeskind09}.

Rotational properties of halo populations can therefore provide clues
as to their origin and evolution, and may allow us to identify
associations sharing a common formation history.  With this in mind,
we develop some simple and flexible distribution functions that can
model rotating stellar populations embedded in dark
halos. Cosmological arguments suggest that dark halos have a universal
Navarro-Frenk-White form (\citealt{nfw}), whilst many stellar
populations in the outer halo are well-approximated by a
power-law. Section 2 provides a flexible set of simple distribution
functions that are powers of energy $E$ and angular momentum $L$ and
which can be adapted to include rotation. We apply the
models to the satellite galaxy, globular cluster and BHB populations
of the Milky Way (Section 3) and M31 (Section 5), with the
implications of our results for the Milky Way given particular attention in Section 4. Finally, Section 6 sums up.

%% file: dfs.tex
\section{Distribution Functions}
\label{sec:dfs}

\subsection{The Even Part of the Distribution Function}

A collisionless system can be described by a phase-space distribution
function (henceforth DF), $F$. The probability that a star occupies
the phase space volume $\mathrm{d}^3x \, \mathrm{d}^3v$ is given by
$F(x,v) \mathrm{d}^3x \, \mathrm{d}^3v$. DFs are a valuable tool for
studying steady-state systems as they replace the impracticality of
following individual orbits with a phase-space probability density
function. The part of the DF that is even in $v_\phi$ fixes the density,
whilst the part that is odd in $v_\phi$ fixes the rotational
properties.

To construct the DFs, we assume a steady-state spherical potential and
a density profile for our halo populations. Of course, these are
tracer populations moving in an external gravitational field generated
by the dark matter halo rather than the self-consistent density
generated through Poisson's equation. For simplicity, we use simple
power-law profiles for the density and potential, namely $\rho \propto
r^{-\alpha}$ and $\Phi \propto r^{-\gamma}$, where $\alpha$ and
$\gamma$ are constants.

\begin{figure}
  \centering
  \includegraphics[width=8cm, height=7cm]{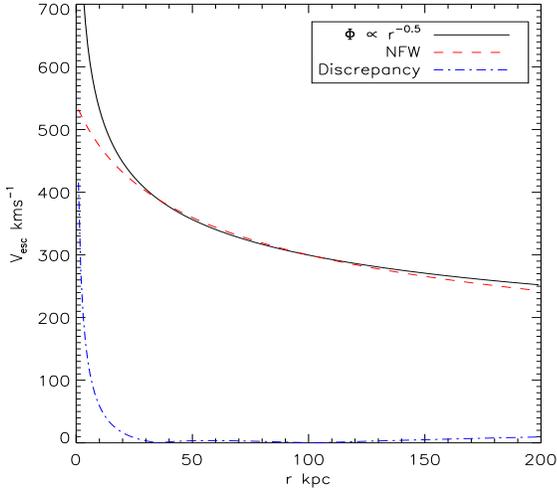}
  \caption[$V_{esc}$]{$V_{\rm esc}$ as a function of radius. The power
    law approximation deviates from a NFW profile at $r <
    20$ kpc. The NFW parameters applicable to the Milky Way
    were taken from \cite{klypin02}. The normalisation of the power
    law is chosen to match the local escape velocity derived in
    \cite{smith07}.}
  \label{fig:vesc}
\end{figure}

The density profiles of dark matter halos in cosmological simulations
resemble the Navarro-Frenk-White or NFW profile
(\citealt{nfw}). Exterior to $r \approx$ 20 kpc, we can approximate
the NFW potential by a power-law profile, as shown in
Fig.~\ref{fig:vesc}. Here, we have used one of the NFW models
applicable to the Milky Way from \citet{klypin02}, and have normalised
the power-law profile to reproduce the local escape speed found by
\cite{smith07}.  We see that a power-law with index $\gamma = 0.5$ is
a good approximation at large radii ($r > 20 \: \mathrm{kpc}$). At small
radii, the approximation breaks down, but in this regime, the bulge
and disk components become important. In any case, our main
application is to tracer populations that reside far out in the halo.
The dark halo mass within a given radius is
\begin{equation}
M(<r)=\frac{\Phi_0 R^{1/2}_\odot}{2\mathrm{G}} r^{1/2}
\end{equation}
where $\Phi_0$ is the normalisation at the solar radius $R_\odot$. For a
virial radius of 250 kpc, the halo mass enclosed is $\approx 8 \times
10^{11} M_\odot$, in good agreement with estimates for the Milky Way
Galaxy and M31 (e.g. \citealt{evans00}; \citealt{xue08};
\citealt{battaglia05}; \citealt{watkins10}).

Previous authors have adopted power-law density profiles for halo
populations with $2 < \alpha < 5$. For example, \cite{harris76},
\cite{zinn85} and \cite{djorgovski94} find a power-law with
$\alpha=3.5$ is a good fit to the Milky Way globular cluster
populations. Classical studies have found $\alpha=3.5$ for stellar
populations in the Milky Way halo~\citep[e.g.][]{freeman87}. More
recently, \cite{bell08} find values of $\alpha$ in the range $2-4$,
although they caution against the use of a single power-law due to the
abundant substructure in the stellar halo. The density profile of the
system of Milky Way satellite galaxies is poorly known due to the small sample size and
incompleteness at large radii. \citet{evans01} find $\alpha=3.4$ is a
good approximation for $r > 20 \: \mathrm{kpc}$ whilst \cite{watkins10}
adopt a power law of $\alpha=2.6$.  Similar profiles have been used
for the M31 halo populations. \cite{crampton85} argue that the radial
profile of the M31 globular cluster system is similar to that of the Milky
Way and so it is reasonable to adopt the same
power-law. \citet{evans00} and \citet{watkins10} use $\alpha$ values
of 3.5 and 2.1 respectively, for the M31 satellites. As a convenient
summary of all this work, we will use a density power-law index of
$\alpha =3.5$, but we discuss the effects of varying $\alpha$ in
Section \ref{applications}.

Armed with these simple forms for the potential and density, we can
give the velocity distribution in terms of the binding energy $E =
\Phi(r)-\frac{1}{2}(v_l^2+v_b^2+v_{\rm los}^2)$ and the total angular
momentum $L = \sqrt{L_x^2+L_y^2+L_z^2}$ as
\begin{equation}
F_{\rm even}(E,L) \propto L^{-2\beta} f(E)
\label{eq:even}
\end{equation}
where
\begin{equation}
\label{eq:df}
f(E) = E^{\frac{\beta(\gamma-2)}{\gamma}+\frac{\alpha}{\gamma}-\frac{3}{2}}
\end{equation}
Here, $\beta$ is the Binney anisotropy parameter (\citealt{binney87}), namely
\begin{equation}
\beta=1-\frac{\langle v^{2}_{\theta} \rangle+ \langle v^{2}_{\phi}
  \rangle}{2 \langle v^2_r \rangle},
\end{equation}
which is constant for the DFs of the form of eqn~(\ref{eq:even}). These DFs are
discussed in greater detail in \cite{evans97}.

\subsection{The Odd Part of the Distribution Function}
So far, our DFs describe non-rotating populations embedded in dark
halos. We now devise an odd part to the distribution function which
generates a one-parameter family of rotating models
\begin{equation}
\label{eq:odd}
F_{\mathrm{odd}}=(1-\eta)\mathrm{tanh}(L_z/\Delta)F_{\mathrm{even}}
\end{equation}
where $\eta$ is a constant. The case $\eta=0$ describes maximum
prograde rotation, whilst $\eta=2$ describes maximum retrograde
rotation. Here, $\Delta$ is a `smoothing' parameter to ease numerical
calculations and soften the Heaviside function.

\begin{figure}
  \centering
  \includegraphics[width=8cm, height=7cm]{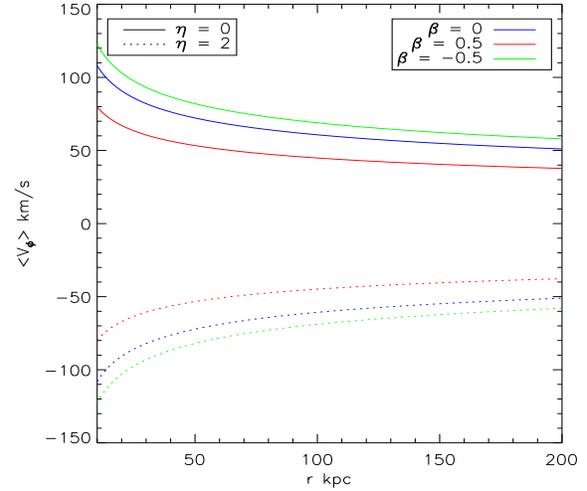}
  \caption[$\langle v_\phi\rangle$]{\small The mean streaming
    velocity $\langle v_\phi\rangle$ as a function of Galactocentric
    distance for $\gamma=0.5$ and $\alpha=3.5$. Solid and dotted lines correspond to maximum prograde
    and retrograde rotation respectively. The profile flattens at
    large radii. Blue, red and
    green lines represent isotropic, radial and tangential velocity
    distributions respectively.}
  \label{fig:vphi}
\end{figure}

To illustrate the effects of the parameter $\eta$, we compute the mean
streaming velocity $\langle v_\phi \rangle$, which is analytic for
$\gamma = 0.5$
\begin{equation}
\label{eq:vphi}
\langle v_\phi \rangle =
\frac{2^{\frac{3}{2}}
  \Gamma(\frac{3}{2}-\beta) \Gamma(2\alpha-4\beta+1)}{\pi
  \Gamma(1-\beta) \Gamma(2\alpha-4\beta + \frac{3}{2})}
(1-\eta) \Phi(r)^{\frac{1}{2}}.
\end{equation}
Table~\ref{tab:vphi} lists $\langle v_\phi \rangle$ values at $r=50,
100\:\mathrm{kpc}$ for three different anisotropy parameters. Fig.~\ref{fig:vphi} shows the radial profile
of the streaming motion for maximum prograde (solid line) and maximum
retrograde (dotted line) distributions.

\begin{table}
\begin{center}
\renewcommand{\tabcolsep}{0.1cm}
\renewcommand{\arraystretch}{0.2}
  \begin{tabular}{  r  c  r  r}
    \hline 
     $\eta$ & $\beta$ & $\langle v_{\phi, \mathrm{50 \, kpc}}\rangle
    \mathrm{kms^{-1}}$ & $\langle v_{\phi, \mathrm{100 \, kpc}}\rangle \mathrm{kms^{-1}}$ \\
    \hline
    0.0 & (-0.5, 0, 0.5) & (82, 72, 53) & (69, 61, 45)\\
    \\
    \\
    0.5 & (-0.5, 0, 0.5)& (41, 36, 27) & (34, 30, 22)\\
    \\
    \\
    1.0 & (-0.5, 0, 0.5) & (0, 0, 0) & (0, 0, 0)\\
    \\
    \\
    1.5 &  (-0.5, 0, 0.5)&  (-41, -36, -27) & (-34, -30, -22)\\
    \\
    \\
    2.0 & (-0.5, 0, 0.5)& (-82, -72, -53) & (-69, -61, -45)\\
    
    \hline
  \end{tabular}
  \caption[$V_\phi$ dependence on $\eta$]
        {\small Typical $\langle v_\phi \rangle$ values evaluated at
          $r=50 \: \mathrm{kpc}$ and $r=100 \: \mathrm{kpc}$ as a function of
          the rotating parameter $\eta$ and the anisotropy parameter
          $\beta$. [$\gamma=0.5, \alpha=3.5$]}
\label{tab:vphi}
\end{center}
\end{table}

\begin{figure*}
  \centering
  \begin{minipage}{0.45\linewidth}
    \centering
  \includegraphics[width=7cm, height=7cm]{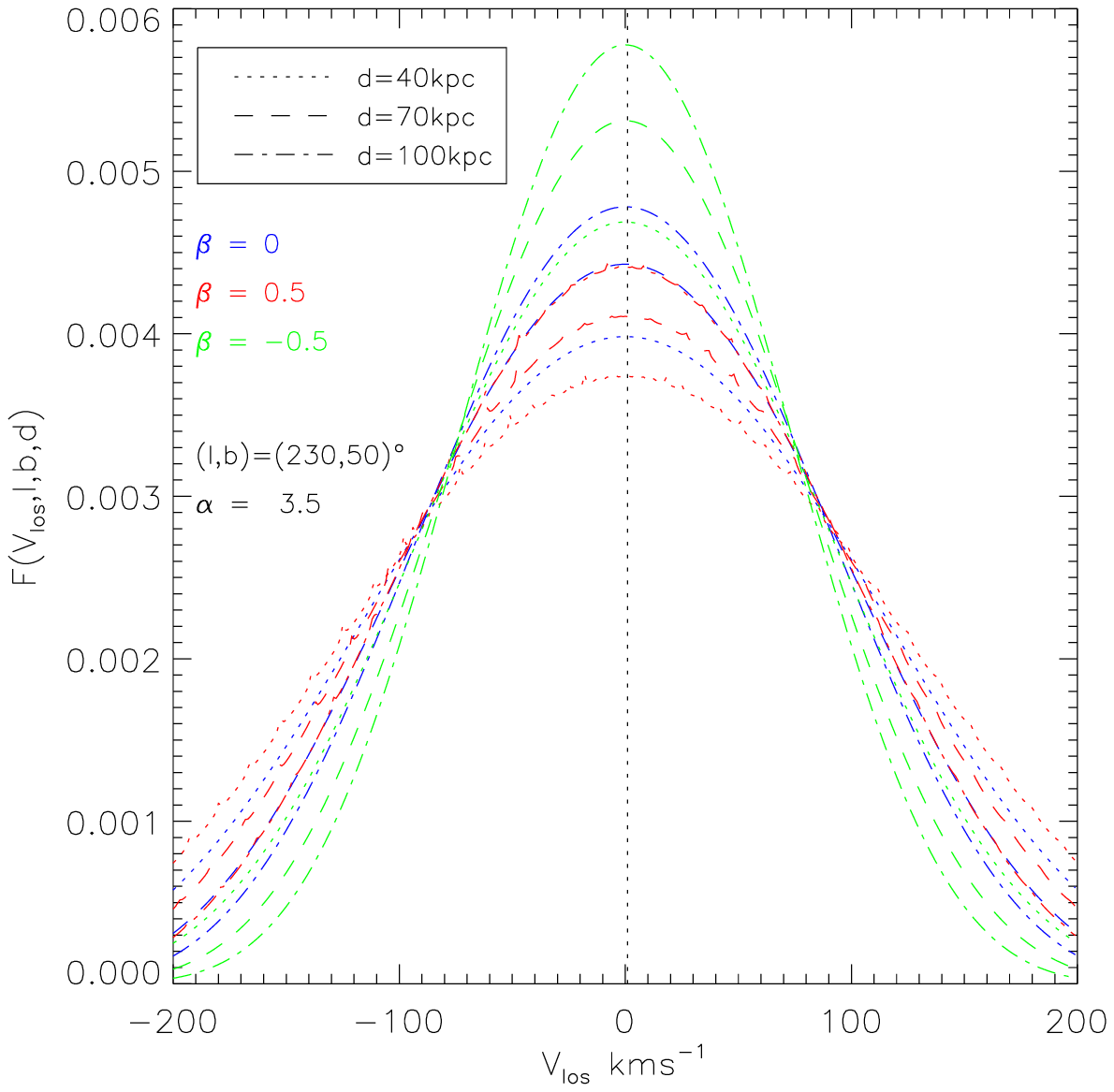}
  \end{minipage}
  \begin{minipage}{0.45\linewidth}
    \centering
  \includegraphics[width=7cm, height=7cm]{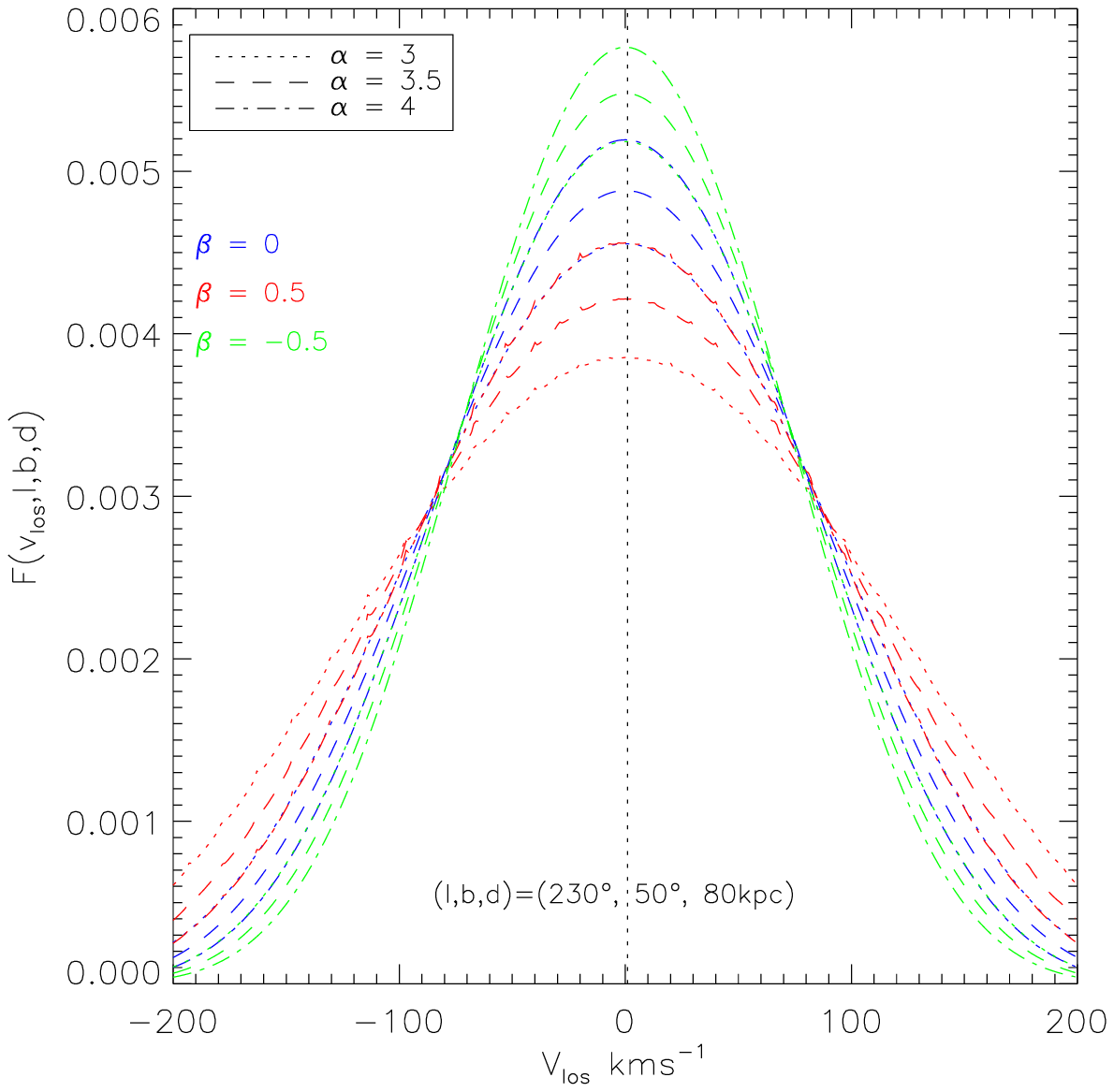}
  \end{minipage}
  \begin{minipage}{0.45\linewidth}
    \centering
  \includegraphics[width=7cm, height=7cm]{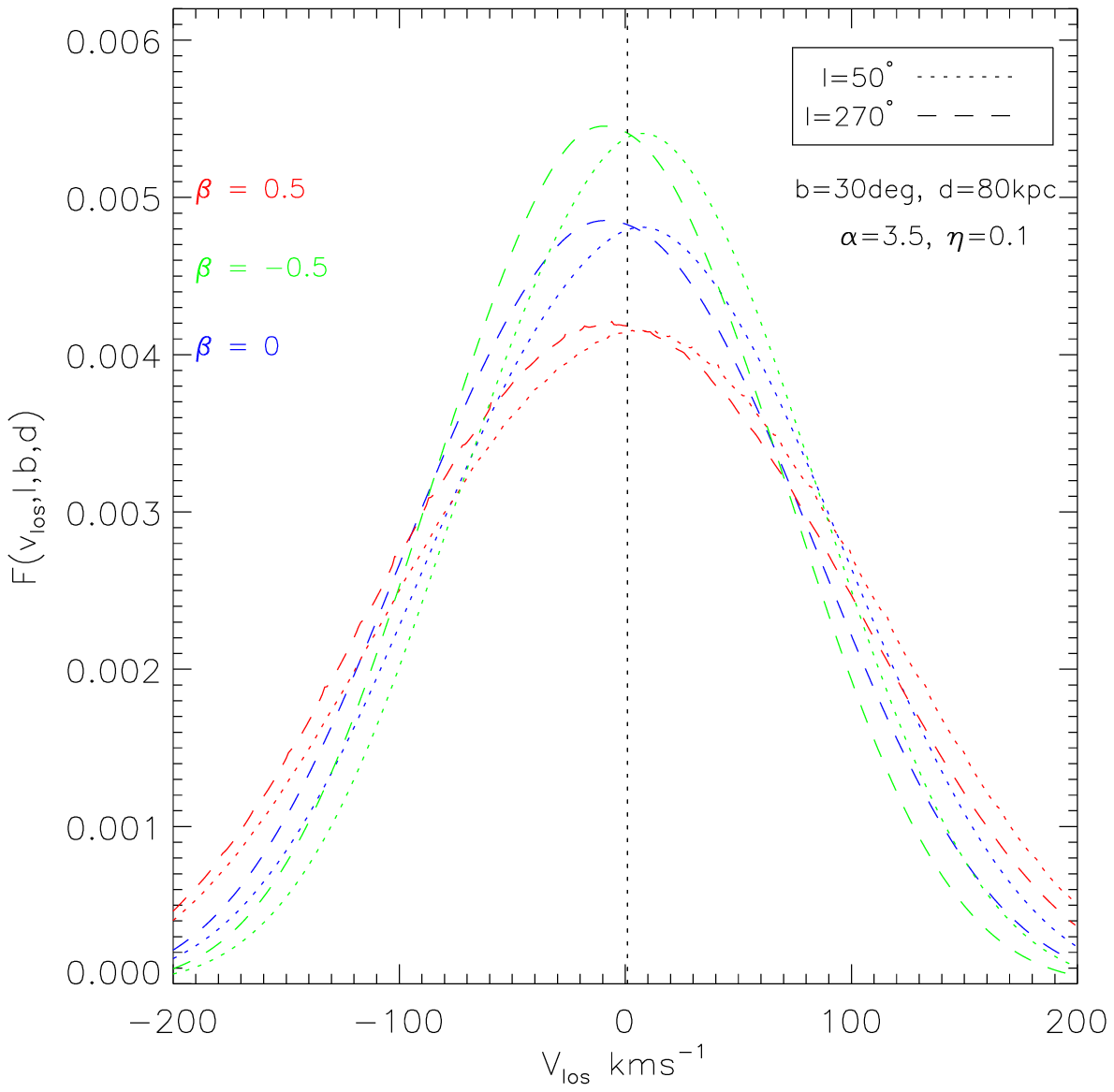}
    \end{minipage}
  \begin{minipage}{0.45\linewidth}
    \centering
  \includegraphics[width=7cm,height=7cm]{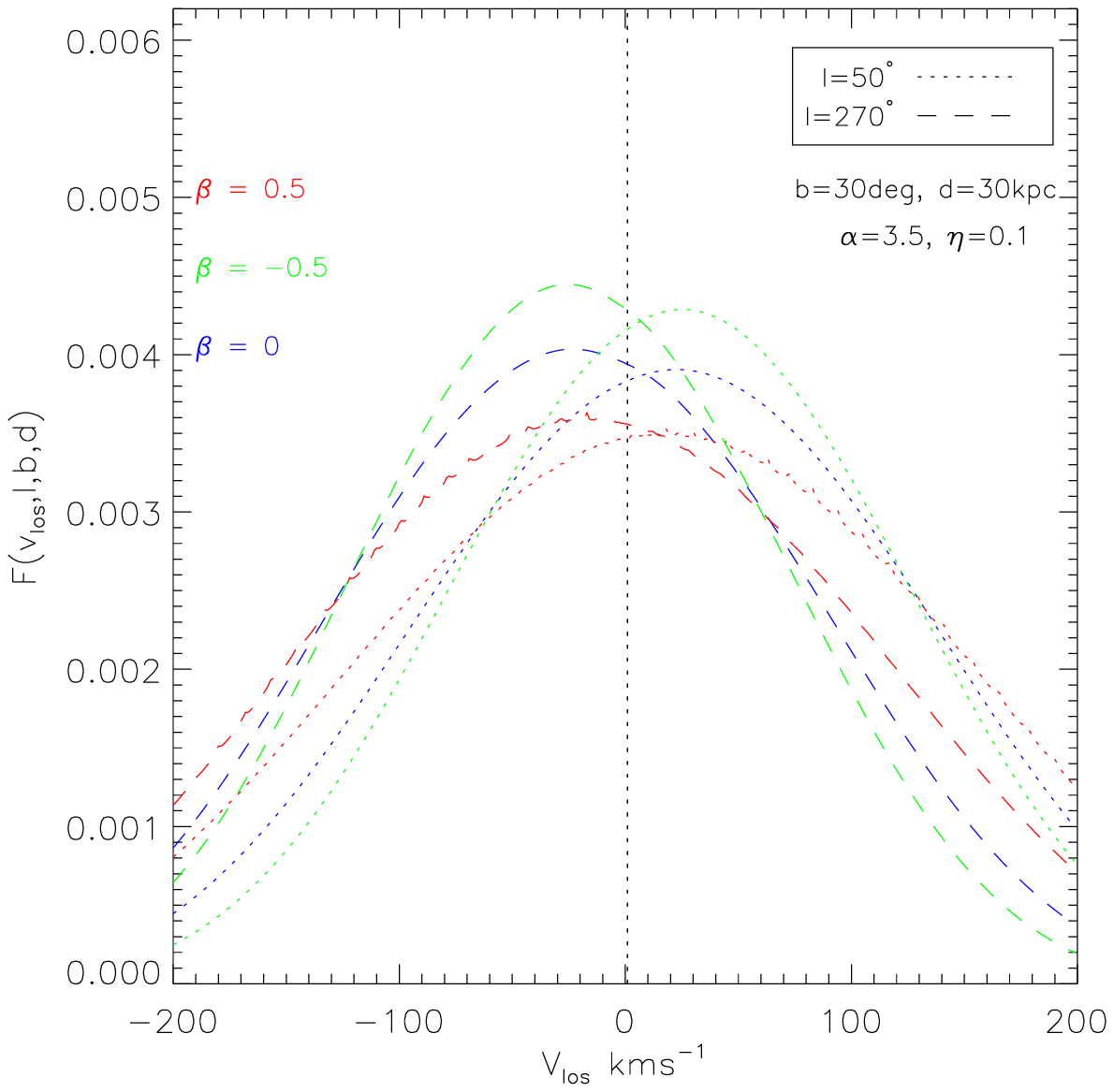}
   \end{minipage}
  \begin{minipage}{0.45\linewidth}
    \centering
  \includegraphics[width=7cm,height=7cm]{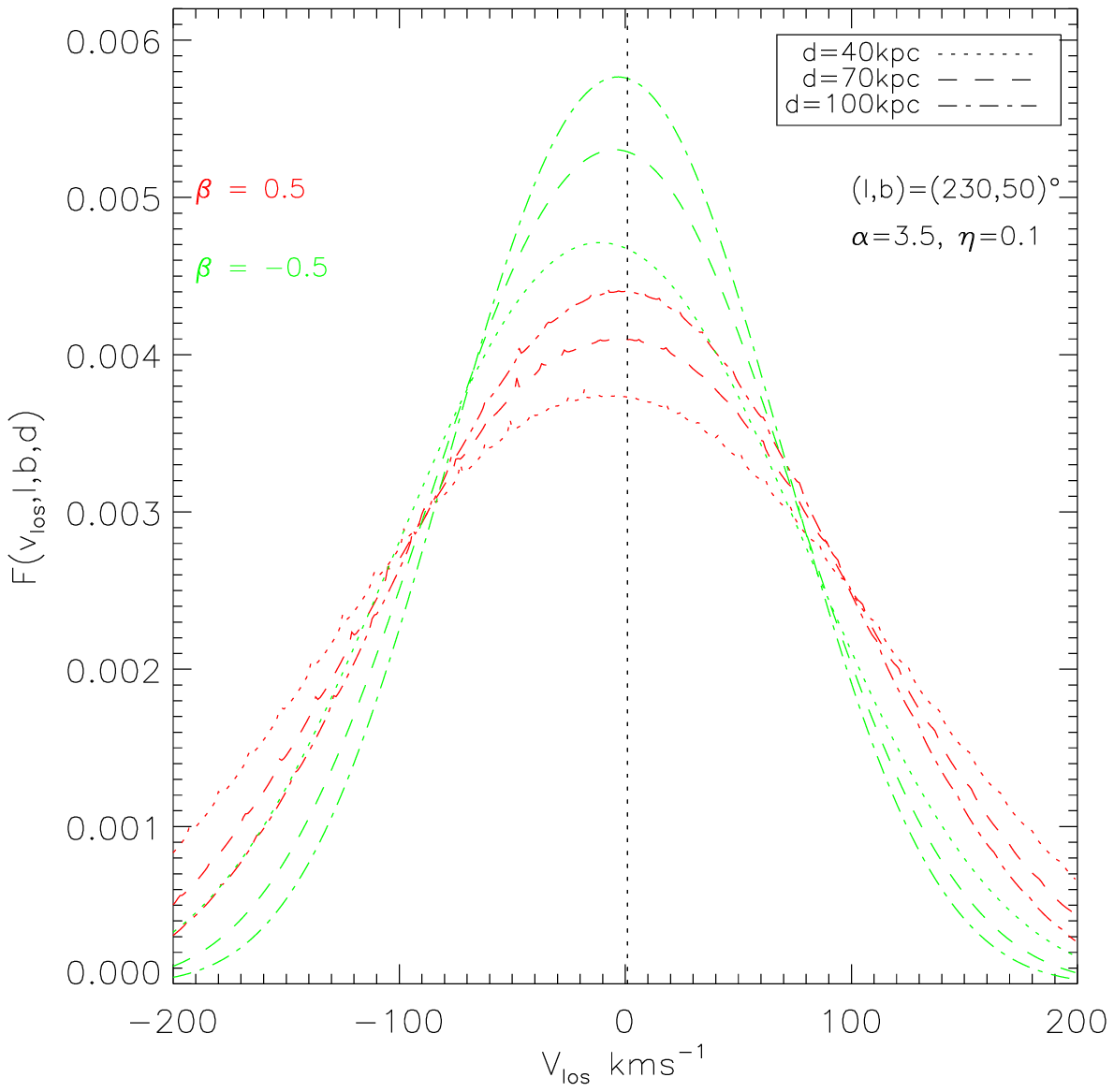}
  \end{minipage}
  \begin{minipage}{0.45\linewidth}
    \centering
  \includegraphics[width=7cm, height=7cm]{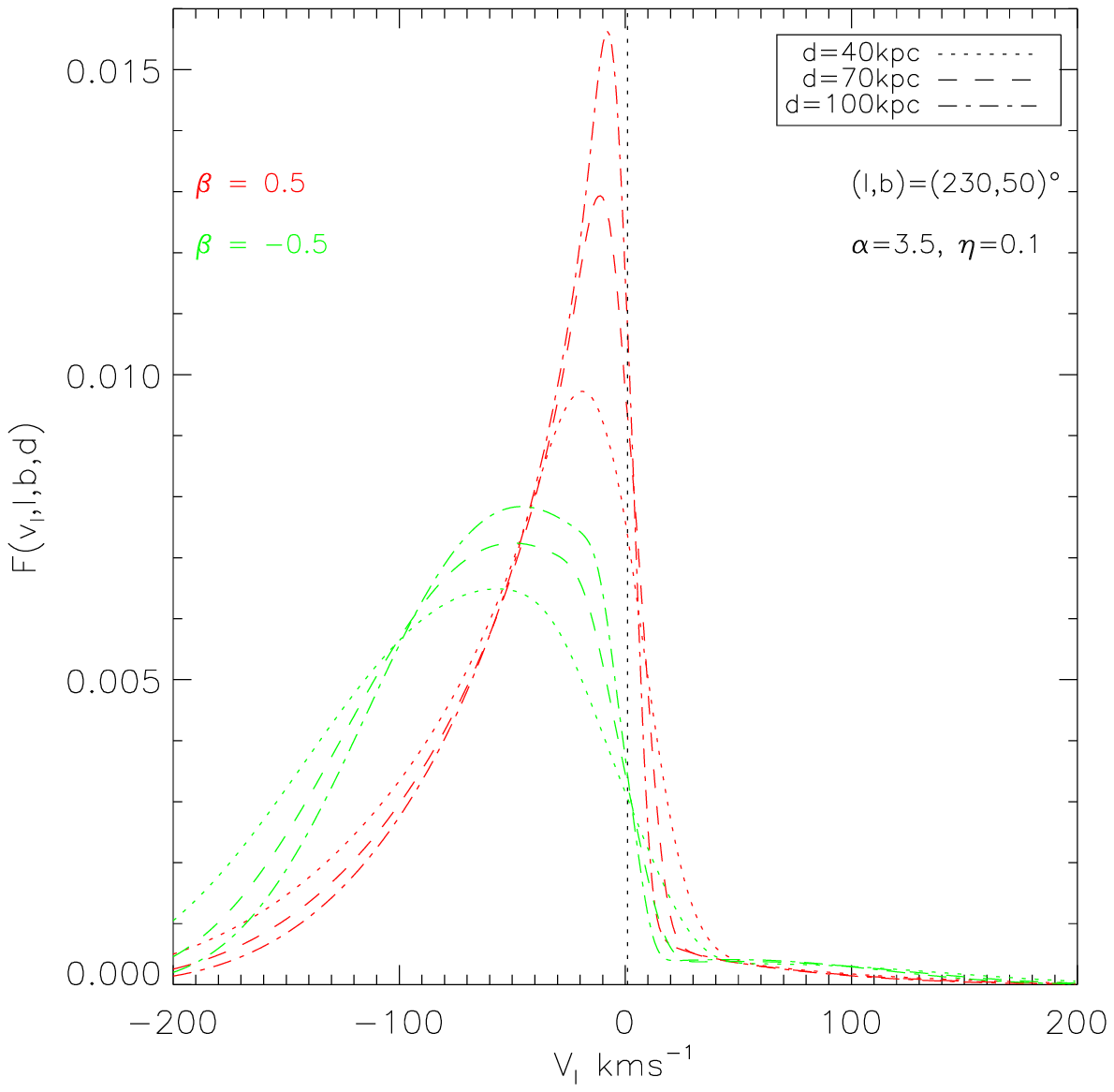}
  \end{minipage}
  \caption[$F(l,b,d,v_{los})$]{Top left: The dependence of the
    distribution function on $v_{\rm los}$ at three different distances in the
    non-rotating case. A random line of sight is chosen (toward the
    Leo constellation) and the density power law is set to
    $\alpha=3.5$. Blue, red and green lines represent isotropic,
    radial and tangential velocity distributions respectively. Top
    right: The distribution function for three different density profile power laws in the
    non-rotating case. The distance is fixed at $d=80 \: \mathrm{kpc}$.  Middle left: The line of sight velocity distribution when rotation is
    included. A prograde rotating case is shown ($\eta=0.1$) for two
    different longitudes. At large distances little difference is seen
    from the non-rotating case. Middle right: As the adjacent
    panel, but now the distance is fixed to $d=30 \: \mathrm{kpc}$. More pronounced
    rotation is seen at smaller distances.  Bottom left: The line
    of sight distribution function for the prograde case at three
    different distances. The same line of sight as the top panels is
    chosen. Very small deviations from a non-rotating distribution are
    seen. The tangential and radial velocity distributions are shown
    by green and red lines respectively.  Bottom right: The
    longitudinal velocity, $v_l$, distribution with the same
    parameters as the adjacent panel. Rotation is much more apparent
    in this case and the distribution is almost entirely skewed to
    negative $v_l$.}
  \label{fig:f_vlos}
\end{figure*}

\subsection{Properties}
\label{props}

It is rare that we possess full six-dimensional phase space
information for any star. However, we can still use a DF by
marginalising over the unknown components. For example, consider a
case where the spatial position and line of sight velocity are well
known, but the proper motions are uncertain.  In this case, we
marginalise over $v_l$ and $v_b$ to obtain the line of sight velocity
distribution (LOSVD):
\begin{equation}
\label{eq:losvd}
F(l,b,d,v_{\rm los})=\int\int F(l,b,d,v_{l},v_{b},v_{\rm los}) \mathrm{d}v_l
\mathrm{d} v_b.
\end{equation}
Once $\gamma$ is fixed as $0.5$, there remain the free parameters, $\beta$, $\alpha$ and $\eta$.

Some properties of the distribution functions are shown in
Fig.~\ref{fig:f_vlos}. The top left panel shows $F(l,b,d,v_{\rm los})$ as a
function of $v_{\rm los}$ at three different distances from the
Sun (for the case $\alpha =3.5$). An arbitrary line of sight toward the
Leo constellation is chosen for illustration,
$(l,b)=(230^\circ,50^\circ)$. Closer
to the Galactic Centre the escape velocity is larger, so the velocity
distributions are broader.  Tangential orbits ($\beta < 0$) have
narrower radial velocity and broader tangential velocity
distributions. The reverse is true for radial orbits ($\beta >
0$).

For distant objects in the Milky Way, $v_{\rm los} \approx v_r$
and so the line of sight velocity distribution reflects the radial
distribution. At small radii, the trend is not so simple and depends
on the particular line of sight in question. The top right panel of
Fig.~\ref{fig:f_vlos} illustrates the dependence of $F(l,b,d,v_{\rm
  los})$ on the power law index of the density profile, $\alpha$. In a
given gravitational potential, more extended tracer populations
(smaller $\alpha$) exhibit larger velocity dispersions than more
centrally-concentrated ones.

The middle panels of Fig.~\ref{fig:f_vlos}
show the changes to the line of sight velocity distributions caused by
the introduction of rotation. At large distances ($d \sim 80 \: \mathrm{kpc}$) the rotating distribution is hardly
distinguishable from the non-rotating case (as $v_{\rm los} \approx v_r$). As expected, radially
biased velocity distributions show the smallest deviation from the
non-rotating case, but even a tangentially biased system may only show
small differences. At smaller distances
($d \sim 30 \: \mathrm{kpc}$) the profile depends on the particular line of
sight as illustrated in the middle right panel.

The bottom panels show the line of sight
velocity distribution and the longitudinal velocity distribution for
prograde rotating systems ($\eta=0.1$). We see that with just the line
of sight velocity, the rotation of a system may be difficult to
constrain.  In particular, if the line of sight component of velocity
is largely due to the radial component, then we can only poorly
constrain rotation. However, applications where the line of sight
velocity has significant contributions from the tangential velocity
components, as for example in the case of M31, can give a more robust
prediction for the rotation. The bottom right panel shows the
distribution as a function of $v_l$. In this case, the distributions
show much more pronounced differences from the non-rotating case. For
this particular line of sight and with $\eta=0.1$, the distribution is
almost entirely skewed to negative $v_l$.
This has important implications for the relevance of accurate proper
motion measurements. The components of motion perpendicular to the
line of sight can give conclusive evidence for rotation, whilst the
deductions from the line of sight velocities rely heavily on the
particular line of sight.

%% file: applications.tex
\section{Applications: The Milky Way Galaxy}
\label{applications}
In this section, we apply our distribution functions to halo
populations in the Milky Way Galaxy.  Observed
heliocentric velocities are converted to Galactocentric ones by
assuming a circular speed of 220 kms$^{-1}$ at the position of the sun
($R_0=8.5$ kpc) with a solar peculiar motion ($U,V,W$)=(11.1,
12.24, 7.25) kms$^{-1}$ (recently updated by
\citealt{schonrich10}). Here, $U$ is directed toward the Galactic
centre, $V$ is positive in the direction of Galactic rotation and $W$
is positive towards the North Galactic Pole.

We set the power law indices for the potential and density to $\gamma=
0.5$ and $\alpha=3.5$ respectively. We
allow the anisotropy $\beta$ and rotation $\eta$ parameters to be
estimated by maximum likelihood analysis of the data. For example, when
we have full distance information and the line of sight velocity, we
construct the likelihood function from the LOSVD:
\begin{equation}
\label{eq:ml}
L(\eta, \beta)=\sum_{i=1}^N \mathrm{log} F(l_i,b_i,d_i,v_{\mathrm{los}_i},
\eta, \beta),
\end{equation}
where $N$ is the number of objects in the population.  Eqn~
(\ref{eq:ml}) gives the two dimensional likelihood as a function of
$\beta$ and $\eta$. As we are mainly interested in constraining
rotation, we sometimes fix $\beta$ and find the likelihood as a
function of $\eta$.

\subsection{Milky Way BHB stars}

\label{sec:bhbs}

\begin{figure*}
  \centering
  \includegraphics[width=11cm,height=10cm]{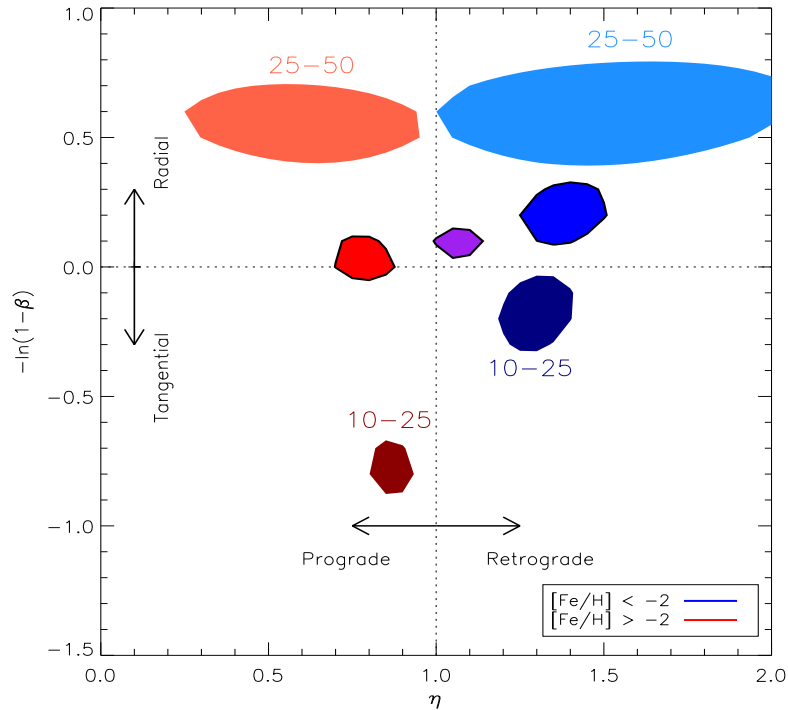}
  \caption[$L(\eta, \beta, r)$ MW BHBs]{\small $1 \sigma$ confidence
    regions for Milky Way halo BHB stars in the $\eta$, $\beta$ plane
    from a maximum
    likelihood analysis. Arrows indicate prograde, retrograde,
    tangential and radial distributions respectively. The overall
    population ($N=3549$ with $r > 10 \: \mathrm{kpc}$) is shown by the
    purple filled contour. Metal poor ($\mathrm{[Fe/H]} < -2$, $N=1135$) and
    metal `rich' ($ \mathrm{[Fe/H]} > -2$, $N=2125$) subsamples are shown by
    blue and red contours respectively. The paler shades show the
    radial bin $25 < r/\mathrm{kpc} < 50$ whilst the darker shades show the
    radial bin $10 < r/\mathrm{kpc} < 25$.}
   
  \label{fig:bhb1}
\end{figure*}

Blue Horizontal Branch (BHB) stars are excellent tracers of halo
dynamics as they are luminous and have an almost constant absolute
magnitude (within a certain colour range). We construct a sample of BHB stars from the
Sloan Digital Sky Survey (SDSS) DR7 release. To select BHBs, we impose constraints on the
(extinction corrected) $u-g$
and $g-r$ colours, the surface gravity ($g_s$) and the effective
temperature ($T_{\mathrm{eff}}$), namely:

\begin{equation}\label{eq:col}
\begin{split}
0.8 < u-g < 1.4 \\
-0.4 < g-r < 0 \\
2 < \mathrm{log}(g_s) < 4 \\ 
7250 < T_{\mathrm{eff}}/ \mathrm{K} < 9700 
\end{split}
\end{equation}

These cuts are similar to those used by previous authors
(e.g. \citealt{yanny00}, \citealt{sirko04a}) and are implemented to
minimise contamination by blue stragglers and main sequence stars. We
only consider stars with $|z| > 4 \: \mathrm{kpc}$, as we wish to exclude disc stars.  \cite{xue08} compiled a sample of 2558 BHB stars from SDSS DR6 using Balmer line profiles to
select suitable candidates. Our larger sample (as we use
the DR7 release) of 5525 stars includes $\approx 88\% $ of the Xue
sample. 

Heliocentric distances are evaluated photometrically by assuming BHBs
have an absolute magnitude of $M_g=0.7$ in the $g$ band \footnote{We check that changing our assumed
absolute magnitude from $M_g=0.7$ to $M_g=0.55$ makes a negligible
difference to our dynamical analysis.}. The
absolute magnitudes of BHB stars are slightly affected by temperature
and metallicity (e.g. \citealt{sirko04a}) but we expect that this variation
makes a negligible difference to our distance estimates. We compare our distance estimates to those of
\cite{xue08}, who take into account such variations,and find agreement to
within $10 \% $. 

We remove contamination of stars belonging to the Sagittarius dwarf
galaxy by masking out tracers lying in the region of the leading and
trailing arms. Our final sample contains 3549 BHB stars (with $r > 10 \:
\mathrm{kpc}$) and so provides a
statistically representative population to study the outer stellar
halo. The heliocentric velocity errors are $\approx 5 \: \mathrm{kms^{-1}}$
on average, at larger radii the error increases due to the decreasing
brightness of the source. The maximum error in the velocity
is $\approx 20 \: \mathrm{kms^{-1}}$.

We use the metallicities derived from the \cite{wbg99} analysis (the
`WBG' method) as this method was adopted specifically for BHB
stars. The method uses the CaII K lines as a metal abundance
indicator and is applicable in the range $-3 < \mathrm{[Fe/H]}
<0$. Only a small fraction ($N \sim 300$) of stars do not have assigned
metallicities or are outside the applicable metallicity range.

Fig.~\ref{fig:bhb1} shows that the \emph{overall}
population (purple contour) has
slight net retrograde rotation ($\eta=1.1$) and has an almost
isotropic distribution. By splitting the sample into metal-poor
($\mathrm{[Fe/H]} < -2$, blue contours) and `metal-rich'
($\mathrm{[Fe/H]} > -2$, red contours) subsamples, we find retrograde and
prograde signals for the two subsets respectively. The mean streaming
motions for the metal-poor and metal-rich samples are $\langle v_\phi\rangle=
-(35 \pm 10) (r/10 \, \mathrm{kpc})^{-1/4} \, \mathrm{kms^{-1}}$ and 
$\langle v_\phi\rangle=
(21 \pm 11) (r/10 \, \mathrm{kpc})^{-1/4} \, \mathrm{kms^{-1}}$
respectively. We remove the radial dependence by averaging over the volume interval ($10 < r/\mathrm{kpc} < 50$) to find $\langle v_\phi\rangle=
-(25 \pm 7) \, \mathrm{kms^{-1}}$ and 
$\langle v_\phi\rangle=
(15 \pm 8) \, \mathrm{kms^{-1}}$ for
the metal-poor and metal-rich populations respectively. These results can be compared to \cite{carollo07} who find a rotating retrograde outer halo and a
weakly prograde rotating inner halo. Our analysis only applies to
the outer halo, but instead of a single population with net retrograde
rotation, we find evidence for two separate populations split by
metallicity (at $\mathrm{[Fe/H]} \sim -2$). In agreement with
\cite{carollo07} (who focus on halo stars within the solar circle) we
find evidence for a net retrograde rotating metal-poor halo component,
but we find that the outer halo also contains a more metal-rich
slightly prograde-rotating component.

We can make a rough estimate of the luminosity associated with these
metal-rich and metal-poor remnants assuming 40\% of the metal-poor sample
($\eta \sim 1.4$, $N_{\mathrm{poor}}=1135$) and 20\% of the metal-rich sample ($\eta
\sim 0.8$, $N_{\mathrm{rich}}=2125$)
are rotating. The fraction of rotating stars can be scaled to the
overall population by calculating the total number of BHBs in a
spherical volume. This requires a normalisation factor for our density
profile ($\rho \propto r^{-3.5}$). We adopt the normalisation applicable to RR
Lyrae stars found by \cite{watkins09} and assume the ratio of BHBs to
RR Lyrae is approximately 2:1. Finally, using photometry for globular clusters
given in \cite{an08}, we find the relation $N_{\mathrm{BHB}}/L \sim
10^{-3} L^{-1}_\odot$ (where $N_{\mathrm{BHB}}$ is found by applying a cut in
colour-colour space - see eqn~\ref{eq:col}).
We find for both the metal-poor and metal-rich rotating
populations a luminosity $L \sim 10^{8}L_{\odot}$. For a mass to
light ratio of $M/L \sim 10$ (appropriate for luminous satellites -
see Fig. 9 in \citealt{mateo98}) this gives a total mass estimate of $M \sim
10^{9}M_{\odot}$. This rough calculation suggests that these rotating
signals could be remnants of two separate accretion events caused by massive satellites with different metallicities and orbital orientations.

\begin{figure}
  \centering
  \includegraphics[width=8cm,height=8cm]{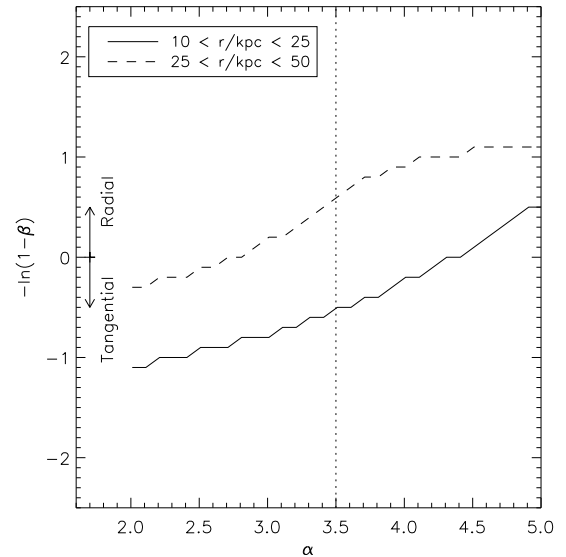}
  \caption[$L(\alpha, \beta, r)$ MW BHBs]{\small The variation in the
    anisotropy parameter $\beta$ with the density profile index
    $\alpha$ for Milky Way halo BHB stars. The solid line and dashed
    lines are for the radial bins, $10 < r/\mathrm{kpc} <
    25$ ($N=2312$) and $25 < r/\mathrm{kpc} < 50$ ($N=1127$) respectively. More
    steeply declining density profiles (larger $\alpha$) give more
    radially anisotropic $\beta$ values. Arrows represent tangential and radial distributions respectively.}
   
  \label{fig:bhb2}
\end{figure}

The different shaded contours in Fig.~\ref{fig:bhb1} show different
radial bins. The paler shades are for the range $25 < r/\mathrm{kpc} < 50$ whilst the
darker shades are for the range $10 < r/\mathrm{kpc} < 25$. Splitting the sample
into radial bins shows the rotation signal is present at all radii
(albeit with less confidence at greater distances due to smaller
numbers) but also shows that the velocity distribution becomes more
radially biased with increasing Galactocentric radii. This result is
in agreement with the findings of cosmological simulations (e.g. \citealt{diemand07}; \citealt{sales07b}). Stars populating the
outer halo are likely remnants of merger events and therefore can be
on eccentric orbits. However, the anisotropy
parameter $\beta$ is dependent on the adopted density profile
(whereas the rotation parameter $\eta$ is independent). More centrally located populations (larger $\alpha$) exhibit
narrower velocity dispersions (see Section \ref{sec:dfs}). To
compensate for the narrower velocity dispersion the unknown tangential
distribution shrinks as we only have a measurement for the line of
sight velocity. Hence, the distribution becomes more radially anisotropic ($\beta$
increases) when we adopt steeper density power laws. This is shown in
Fig.~\ref{fig:bhb2} for two radial bins. To break this degeneracy we require a full analysis of the density
profile of the BHB sample. As our main result regarding rotation is unaffected we defer this task to a future paper.

\begin{figure}
  \centering
  \includegraphics[width=8.5cm,height=8.5cm]{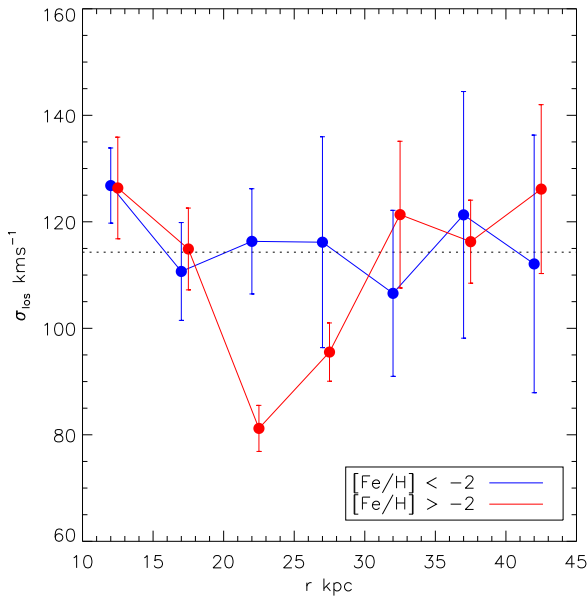}
  \caption[$\sigma$ MW BHBs]{\small Line of sight velocity dispersion as a function of
    Galactocentric radii for Milky Way halo BHB stars. The metal poor ($[\mathrm{Fe/H}] < -2$) and metal
    rich ($[\mathrm{Fe/H}] > -2$) components are given by the blue and
  red lines respectively. An obvious cold feature is apparent in the
  metal rich population at $r \sim 25 \: \mathrm{kpc}$. The dotted line
  is the average $\sigma_{\mathrm{los}}$ for the whole population.}
   
  \label{fig:bhb3}
\end{figure}

Evidence for substructure in the metal-rich sample is
shown in Fig.~\ref{fig:bhb3}. Here, we show the line of sight
velocity dispersion, $\sigma_{\mathrm{los}}$, as a function of Galactocentric
radii. We formulate error bars by assuming Gaussian distributed
errors and apply a maximum likelihood routine to find the
$1 \sigma$ deviations. There is an obvious `cold' structure in the metal-rich sample for $ 20
< r/\mathrm{kpc} < 30 $. This could be evidence for a shell type structure
caused by a relatively recent accretion event. The lack of such a feature in the
metal-poor sample suggests that the net retrograde signal is not a consequence of a single accretion event. The dichotomous nature
of the outer halo may reflect two populations with very different
origins. This is discussed further in Section \ref{sec:discussion}.

We note that a larger sample of BHB stars extending to larger
Galactocentric distances, will be invaluable in studying the outer
reaches of the stellar halo. To date very few tracers are available,
but future spectroscopic surveys are likely to increase the sample of
distant BHBs with measured velocities. For example the proposed LSST
mission, will provide a vast number of spectroscopic targets
out to a magnitude limit of $r \sim 27$.

\subsection{Milky Way globular clusters}
\begin{table*}
\begin{center}
\renewcommand{\tabcolsep}{0.1cm}
\renewcommand{\arraystretch}{0.2}
\begin{tabular}{  l  r  r  r  r  r  r  c}
    \hline 
     Name & r $ (\mathrm{kpc})$ & [Fe/H] & $v_{\mathrm{h}}$
     (kms$^{-1})$ &  $\mu_{\alpha} \, \mathrm{cos}(\delta)$
    (mas yr$^{-1}$) &  $\mu_{\delta}$ (mas yr$^{-1}$) & $v_\phi$
     (kms$^{-1})$ & Ref\\
    \hline
    NGC 288 & 11.8 & -1.24 & -46.6 & $4.40 \pm 0.23$ & $-5.62 \pm
    0.23$ & -27 $\pm$ 18 & 1,2
    \\
    \\
    NGC 1851 & 17.2 & -1.26 & 320.9 & $1.28 \pm 0.68$ & $2.39 \pm
    0.65$ & 134 $\pm$ 29 & 1,2
    \\
    \\
    NGC 1904 & 18.9 & -1.54 & 207.5 & $2.12 \pm 0.64$ & $-0.02 \pm
    0.64$ & 83 $\pm$ 29 & 1,2
    \\
    \\
    NGC 2298 & 16.0 & -1.85 & 148.9 & $4.05 \pm 1.00$ & $-1.72 \pm
    0.98$ & -27 $\pm$ 30 &1,2
    \\
    \\
    NGC 2808 & 11.2 & -1.37 & 93.6 & $0.58 \pm 0.45$ & $2.06 \pm 0.46$
    & 82 $\pm$ 16 & 1,3
    \\
    \\
    NGC 4147 & 21.1 & -1.83 & 183.2 & $-1.85 \pm 0.82$ & $-1.30 \pm
    0.82$ & 67 $\pm$ 65 & 1,2
    \\
    \\ 
    NGC 4590 & 10.3 & -2.06 & -95.2 & $-3.76 \pm 0.66$ & $1.79 \pm
    0.62$ & 294 $\pm$ 30 & 1,2
    \\
    \\ 
    NGC 5024 & 19.0 & -2.07 & -79.1 & $0.50 \pm 1.00$ & $-0.10 \pm
    1.00$ & 240 $\pm$ 85 & 1,2
    \\
    \\
    NGC 5272 & 12.1 & -1.57 & -148.6 & $-1.10 \pm 0.51$ & $-2.30 \pm
    0.54$ & 105 $\pm$ 24 & 1,2
    \\
    \\
    NGC 5466 & 17.0 & -2.22 & 107.7 & $-4.65 \pm 0.82$ & $0.80 \pm
    0.82$ & -63 $\pm$ 64 & 1,2
    \\
    \\
    NGC 6934 & 12.4 & -1.54 & -411.4 & $1.20 \pm 1.00$ & $-5.10 \pm
    1.00$ & -67 $\pm$ 60 & 1,2
    \\
    \\    
    NGC 7078 & 10.5 & -2.22 & -107.5 & $-0.95 \pm 0.51$ & $-5.63 \pm
    0.50$ &  129 $\pm$ 25 & 1,2
    \\
    \\
    NGC 7089 & 10.4 & -1.62 & -5.3 & $5.90 \pm 0.86$ & $-4.95 \pm
    0.86$ & -84 $\pm$ 41 & 1,2
    \\
    \\
    Pal 3 & 92.9 & -1.66 & 83.4 & $0.33 \pm 0.23$ & $0.30 \pm 0.31$ &
    146 $\pm$ 95 & 1,2
    \\
    \\
    Pal 5 & 17.8 & -1.38 & -55.0 & $-1.78 \pm 0.17$ & $-2.32 \pm 0.23$
    & 42 $\pm$ 34 & 1,2
    \\
     \hline
  \end{tabular}
  \caption[MW globular clusters]{ The Milky Way halo globular cluster
    sample with available proper motions. We give the Galactocentric
    radii $r$ (note we correct the \cite{harris96} values to a solar
    position of $R_\odot=8.5 \, \mathrm{kpc}$), the metallicity [Fe/H], the
    heliocentric velocity, $v_{\rm h}$, the heliocentric rest frame
    proper motions in right ascension and declination and
    the rotational velocity, $v_\phi$. References: (1)
    \cite{harris96}, (2) \cite{dinescu99}, (3) \cite{dinescu07}.}
\label{tab:mw_gc}
\end{center}
\end{table*}

\begin{figure*}
  \centering
  \begin{minipage}{0.45 \linewidth}
    \centering
    \includegraphics[width=8cm,height=8cm]{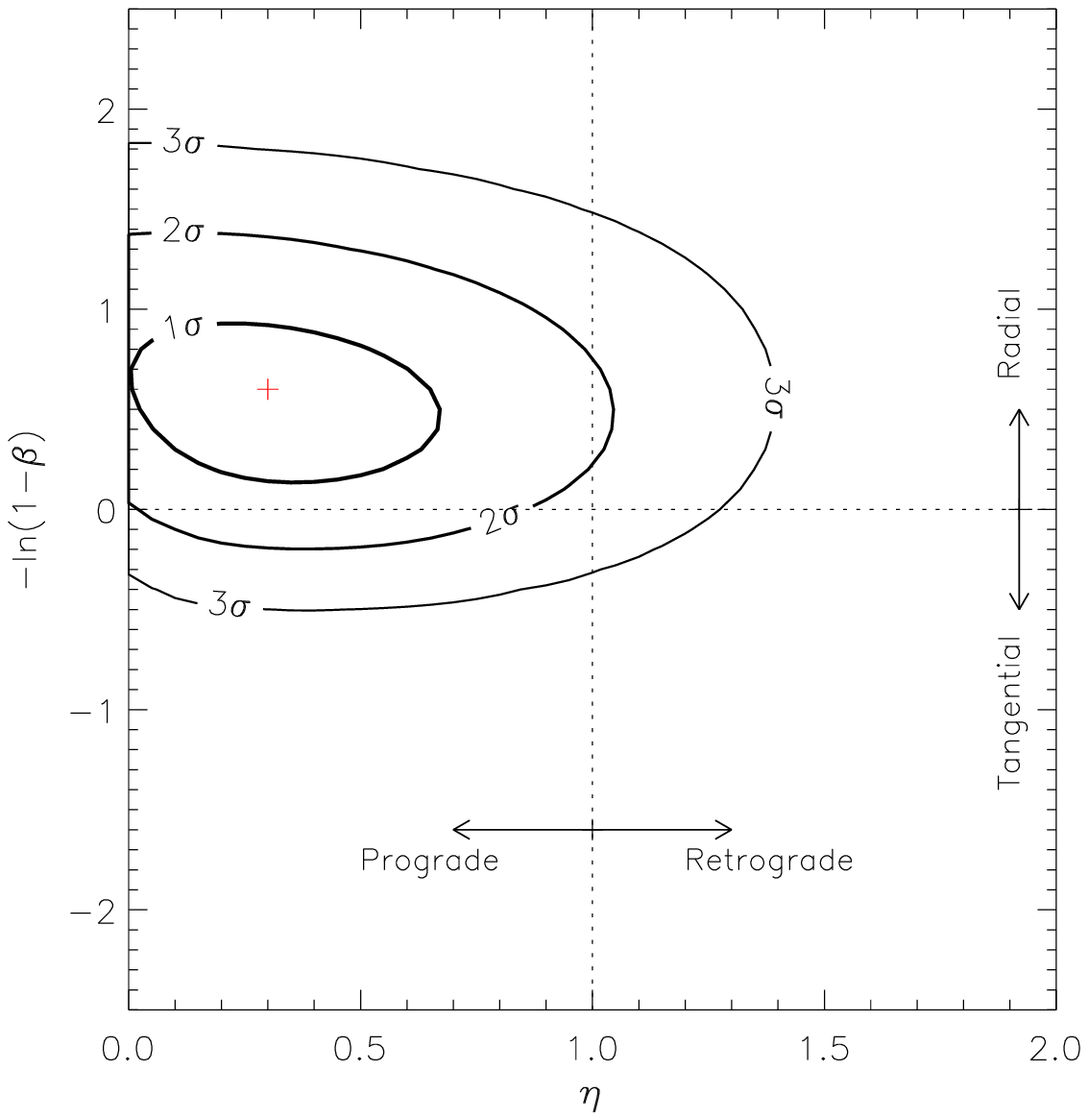}
  \end{minipage}
  \begin{minipage}{0.45 \linewidth}
    \centering
    \includegraphics[width=8cm,height=8cm]{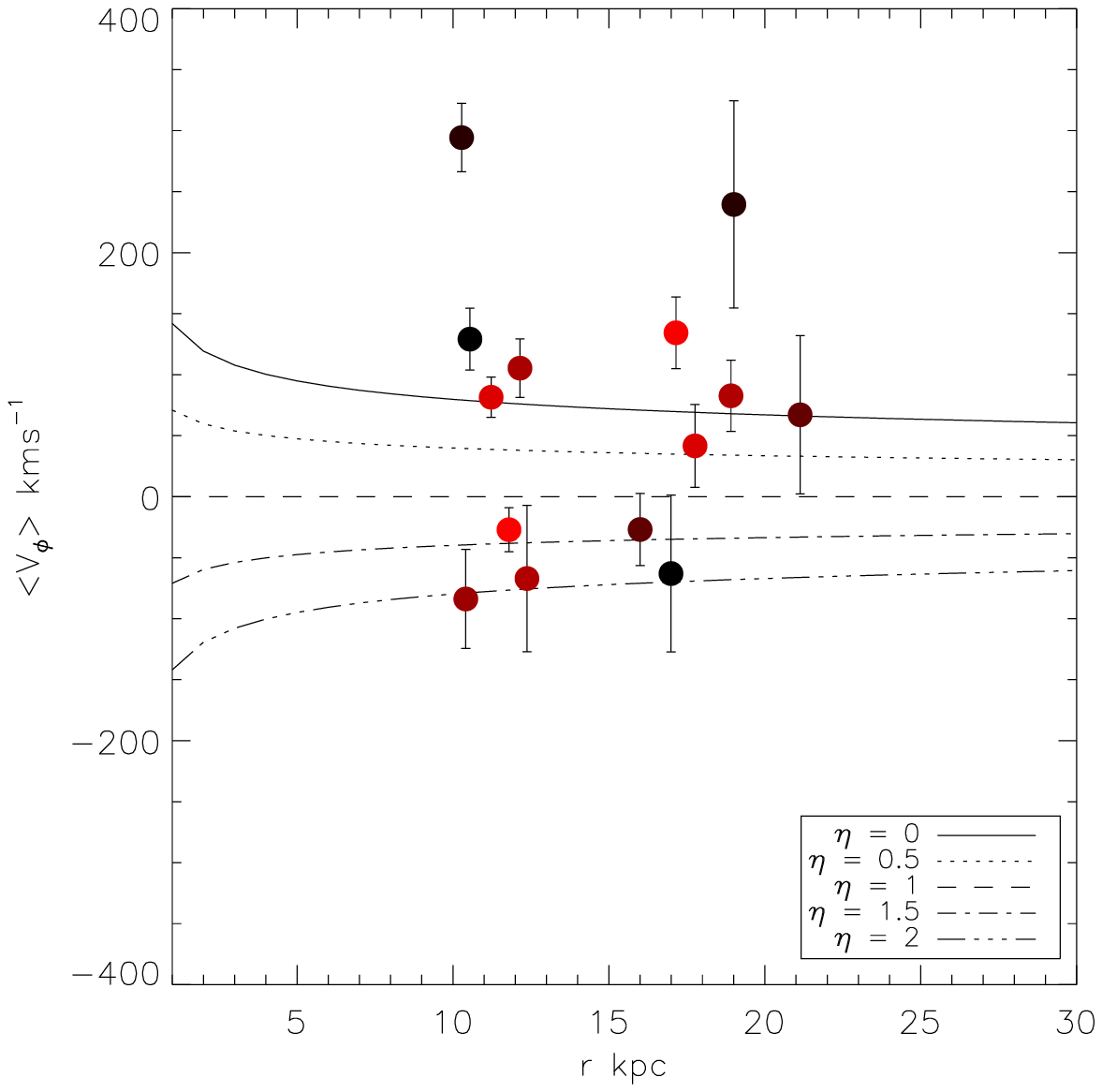}
    \end{minipage}
    \caption[MW GCs]{\small Left: The confidence
      contours for Milky Way halo globular clusters in the $\eta$, $\beta$ plane evaluated from a 2D
      Maximum likelihood analysis. Arrows illustrate
      prograde, retrograde, radial and tangential distributions
      respectively. We use $-\mathrm{ln}(1-\beta)$ on the y-axis to
      make the ranges occupied by radial and tangential models
      symmetric. The Milky Way globular clusters with
      available proper motions ($N=15$) show evidence for prograde rotation
      ($\eta \sim 0.3$) and a mildly radial velocity distribution
      ($\beta \sim 0.5$). Right: The streaming motion as a function of
      Galactocentric radius for a mildly radial distribution $\beta=0.5$
      (see Fig.~\ref{fig:vphi}). Different linestyles represent
      different rotation parameters ($\eta$). Overplotted are the
      $v_\phi$ velocity components of the Milky Way globular clusters with
      available proper motions. The points are colour coded according
      to metallicity, black being the most metal-poor
      ($\mathrm{[Fe/H]} \sim -2.2$) and red being the most metal-rich
      ($\mathrm{[Fe/H]} \sim -1.2$). There is no obvious correlation
      with metallicity but the two clusters with particularly large positive
      $v_\phi$ ($\sim 200-300 \, \mathrm{kms^{-1}}$), NGC 4590 and NGC
      5024, are also metal poor ($\mathrm{[Fe/H]} < -2$). Note that $\sim 5$ of the globular
      clusters have negative $v_\phi$, but the majority have positive
      $v_\phi$. As there are only 15 objects in the sample the net
      prograde motion may not be representative of the whole
      population. Pal 3 is not shown as its distance of $r \sim
      90 \, \mathrm{kpc}$ is beyond the range of the plot, but note
      that it too has a positive $v_\phi$ (albeit with large formal
      uncertainties, $146 \pm 95 \, \mathrm{kms^{-1}}$)}
  
  \label{fig:mw_gc_pm}
\end{figure*}

The kinematics of the Milky Way globular clusters have been studied by
a number of authors (e.g. \citealt{frenk80}; \citealt{zinn85}). We
apply the methods adopted in the previous section to model the halo
globular clusters. Using the data taken from \cite{harris96}, we
restrict attention to those at galactocentric distances $r > 10 \:
\mathrm{kpc}$.

With line of sight velocities alone, our sample of 41 globular
clusters poorly constrains $\beta$ and $\eta$. We improve this
analysis using the proper motion measurements from \cite{dinescu99}
and \cite{dinescu07},
which include 15 of our halo clusters (see Table \ref{tab:mw_gc}). We use a similar procedure to
\cite{wilkinson99} and convolve our probabilities with an error
function. Eqn~(\ref{eq:losvd}) becomes
\begin{eqnarray}
F(l,b,d,v_{los}) &=& \int\int E_1(v_l)E_1(v_b) \\
                 && \times \:  F(l,b,d,v_{l},v_{b},v_{los})
\mathrm{d}v_l \mathrm{d}v_b, \notag
\end{eqnarray}
where $E_1(v)$ is the error function. We assume the Lorentzian given by
\begin{equation}
E_1(v)=\frac{1}{\sqrt{2}\pi\sigma_1}\frac{2\sigma^2_1}{2\sigma_1^2+(v-v_{\mathrm{obs}})^2},
\end{equation} 
where $\sigma_1$ is related to the published error estimate by
$\sigma_1=0.477 \sigma_{\mathrm{meas}}$. The properties of such
functions are discussed further in Appendix A of
\cite{wilkinson99}.
We show in the left hand panel of Fig.~\ref{fig:mw_gc_pm} that the halo clusters have a mildly radial
distribution ($\beta = 0.5^{+0.1}_{-0.3}$) and net prograde rotation
($\eta =0.3^{+0.35}_{-0.3}$). We find a mean streaming motion of
$\langle v_\phi \rangle= (58 \pm 27) (r/ \,10\mathrm{kpc})^{-1/4} \,
\mathrm{kms^{-1}}$ which evaluates to $\langle v_\phi \rangle \sim
(60-50)\, \mathrm{kms^{-1}}$ in the range $r=(10-20) \, \mathrm{kpc}$. This agrees with previous
authors (e.g. \citealt{frenk80}; \citealt{zinn85}) who have found that
the halo globular clusters are primarily pressure supported, but show
weak rotation with $v_{\mathrm{rot}} \sim 50-60 \, \mathrm{kms^{-1}}$. 

The right panel of Fig.~\ref{fig:mw_gc_pm} shows the $v_\phi$ velocity
components of the globular clusters overplotted on a mildly radial
model for $\langle v_\phi(r) \rangle$. The majority of the globular
clusters have positive $v_\phi$ and hence the net streaming motion is
prograde. However, 5 of the halo globular clusters have negative
$v_\phi$. As our sample only consists of 15 objects, we cannot conclude
that this net prograde rotation is ubiquitous for the whole
population. There is no obvious correlation between rotational
velocity and metallicity, suggesting independent accretion events
rather than dissipative collapse as the formation mechanism.

The majority of the halo globular clusters in our sample have
metallicities  between
$-2 < [\mathrm{Fe/H}] < -1$ and Galactocentric radii in the range
$10 < r/\mathrm{kpc} < 20$ (with the notable exception of Pal
3). It is interesting that these properties are in common with the relatively
metal-rich halo BHB stars, which also exhibit net prograde rotation
(albeit a weaker signal). We suggest that some of the globular clusters with
positive $v_\phi$ may share a common accretion history with these
relatively metal-rich field halo stars.

\subsection{Milky Way Satellites}
\begin{table}
\begin{center}
\renewcommand{\tabcolsep}{0.1cm}
\renewcommand{\arraystretch}{0.2}
  \begin{tabular}{  l  r  r  r  r  c}
    \hline 
     Name & $l \, (^\circ)$ & $b \, (^\circ)$ & $d \, (\mathrm{kpc})$ &
    $v_{\mathrm{h}}$ (kms$^{-1})$ & Ref\\
    \hline
    Boo I & 358.1 & 69.6 & 62.0 & 99.0  & 1 
    \\
    \\
    Boo II & 353.7 & 68.9 & 46.0 & -117.0  & 2 
    \\
    \\
    Can Ven I & 74.3 & 79.8 & 224.0 & 30.9  &  3
    \\
    \\
    Can Ven II & 113.6 & 82.7 & 151.0 & -128.0 & 3 
    \\
    \\
    Coma & 241.9 & 83.6 & 44.0 & 98.1  & 3 
    \\
    \\
    Hercules & 28.7 & 36.9 & 138.0 & 45.0  & 3 
    \\
    \\
    Leo IV & 265.4 & 56.5 & 158.0 & 132.3 & 3 
    \\
    \\
    Leo  V & 261.9 & 58.5 & 180.0 & 173.3  & 4 
    \\
    \\
    Segue I & 220.5 & 50.4 & 23.0 & 206.0  & 5 
    \\
    \\
    Segue II & 149.4 & -38.1 & 35.0 & -39.2  & 6 
    \\
    \\
    Ursa Maj II & 152.5 & 37.4 & 32.0 & -116.5  & 3 
    \\
    \\
    Willman I & 158.6 & 56.8 & 38.0 & -12.3  & 1 
    \\
        
    \hline
  \end{tabular}
  \caption[Ultra-Faint MW Satellites]{ The properties of the
    ultra-faint Milky Way satellites. We give the Galactic coordinates $(l,b)$,
    the heliocentric distance $d$, and the heliocentric velocity
    $v_{\rm h}$. 
    References: (1) \citealt{martin07}, (2) \citealt{koch09},
          (3) \citealt{simon07}, (4) \citealt{belokurov08}, (5)
    \citealt{geha09}, (6) \citealt{belokurov09}}
\label{tab:mw_uf}
\end{center}
\end{table}

\begin{figure}
  \centering
  \includegraphics[width=8cm, height=8cm]{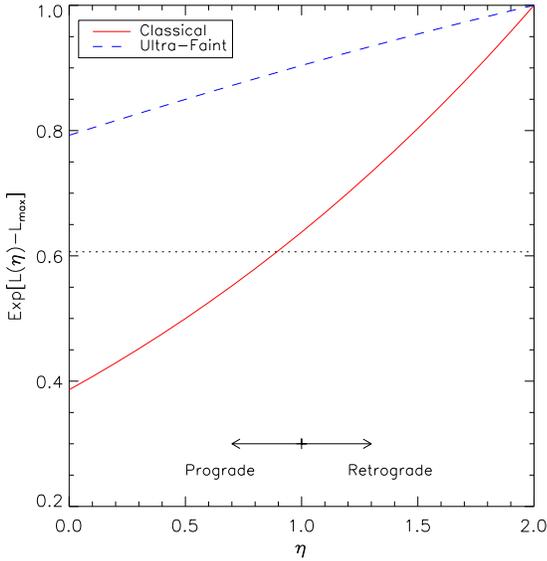}
  \caption[$L(/eta)$ MW satellites]{\small Maximum likelihood analysis
    of the rotation parameter, $\eta$, for Milky Way
    satellites. The velocity anisotropy is kept fixed at $\beta=0$. The
    y-axis, $\mathrm{Exp}[L(\eta)-L_{\rm max}]$, is the reduction of the
    likelihood for a particular $\eta$ value with respect to the
    maximum likelihood. The red (solid) and blue (dashed) lines
    represent the classical and ultra-faint satellites
    respectively. Rotation is poorly constrained in both cases.}
  \label{fig:mw_sat}
\end{figure}

\begin{table*}
\begin{center}
\renewcommand{\tabcolsep}{0.1cm}
\renewcommand{\arraystretch}{0.2}
  \begin{tabular}{  l  r  r  r  r  r  r r  c  }
    \hline 
     \small Name & $l \, (^\circ)$ & $b \, (^\circ)$ & $d \,(\mathrm{kpc})$ &
    $v_{\mathrm{h}}$ ( kms$^{-1}$) &  $\mu_{\alpha}$
    (mas cent$^{-1}$) &  $\mu_{\delta}$ (mas cent$^{-1}$) & $v_\phi$
     (kms$^{-1})$ & Ref\\
    \hline
    Carina & 260.0 & -22.2 & 101.0 & 224.0  & 22 $\pm$ 9 &
    15 $\pm$ 9 & 68 $\pm$ 43 & 1,2 \\
    \\
    \\
    Fornax & 237.1 & -65.7 & 138.0 & 53.0  & 48 $\pm$ 5 &
    -36 $\pm$ 4 & -76 $\pm$ 27 & 1,3\\
    \\
    \\
    Draco & 86.4 & 34.7 & 82.0 & -293.0 & 19 $\pm$ 13 &
    -3 $\pm$ 12 & 17 $\pm$ 47 & 1,4 \\
    \\
    \\
    Leo I & 226.0 & 49.1 & 250.0 & 286.0 & - & - & - & 1\\
    \\
    \\
    Leo II & 220.2 & 67.2 & 205.0 & 76.0  & - & - & - & 1\\
    \\
    \\
    LMC & 280.5 & -32.9 & 50.0 & 278.0 & 196 $\pm$ 4 &
    44 $\pm$ 4 & 82 $\pm$ 8 & 5,6 \\
    \\
    \\
    Sagittarius & 5.6 & -14.1 & 24.0 & 140.0 & -283 $\pm$ 20 &
    -133 $\pm$ 20 & 49 $\pm$ 22 & 1,7 \\ 
    \\
    \\
    Sculptor & 287.5 & -83.2 & 79.0 & 108.0 & 9 $\pm$ 13 &
    2 $\pm$ 13 & 126 $\pm$ 49 & 1,8\\
    \\
    \\
    Sextans & 243.5 & 42.3 & 86.0 & 227.0 & -26 $\pm$ 41 &
    10 $\pm$ 44 & 193 $\pm$ 153 & 1,9 \\
    \\
    \\
    SMC & 302.8 & -44.3 & 60.0 & 158.0 & 75 $\pm$ 6 &
    -125 $\pm$ 6 & 44 $\pm$ 16 & 5,6 \\
    \\
    \\
    Ursa Min & 105.0 & 44.8 & 66.0 & -248.0  & -50 $\pm$ 17 &
    22 $\pm$ 16 & -87 $\pm$ 48 & 1,10 \\
    
    \hline
  \end{tabular}
  \caption[Classical MW Satellites]
  {The properties of the classical Milky Way satellites. Here, we give
    the Galactic coordinates $(l,b)$, the heliocentric distance $d$,
    the heliocentric velocity $v_{\rm h}$, and the heliocentric rest frame proper motions in
    right ascension and declination and
    the rotational velocity, $v_\phi$.  References: (1)
    \citealt{mateo98}, (2) \citealt{piatek03}, (3) \citealt{piatek07},
    (4) \citealt{piatek08a}, (5) \citealt{kara04},
    (6) \citealt{piatek08b}, (7) \citealt{dinescu05}, (8)
    \citealt{piatek06}, (9) \citealt{walker08}, (10) \citealt{piatek05}}
\label{tab:mw_c}
\end{center}
\end{table*}

Our final application focusing on our own Milky Way Galaxy looks at the
satellite galaxies. Fig.~\ref{fig:mw_sat} shows the result of the maximum
likelihood analysis keeping the velocity anisotropy fixed at $\beta=0$. We have split the satellites into classical ($L
\sim 10^6 L_\odot$) and ultra-faint ($L \sim 10^2-10^4 L_\odot$)
samples (see Tables \ref{tab:mw_c} and \ref{tab:mw_uf}). There is some
controversy as to whether these groups are separate populations or
just the bright and faint components of a single population. We can
see from Fig.~\ref{fig:mw_sat} that the rotation is poorly
constrained. The $1\sigma$ confidence limit (horizontal dotted line)
encompasses non-rotating, retrograde and prograde solutions. We
illustrate the isotropic case ($\beta=0$) in Fig.~\ref{fig:mw_sat} but
the same form is seen for radial or tangential distributions.

\begin{figure*}
  \centering
  \begin{minipage}{0.55 \linewidth}
    \centering
  \includegraphics[width=9cm, height=9cm]{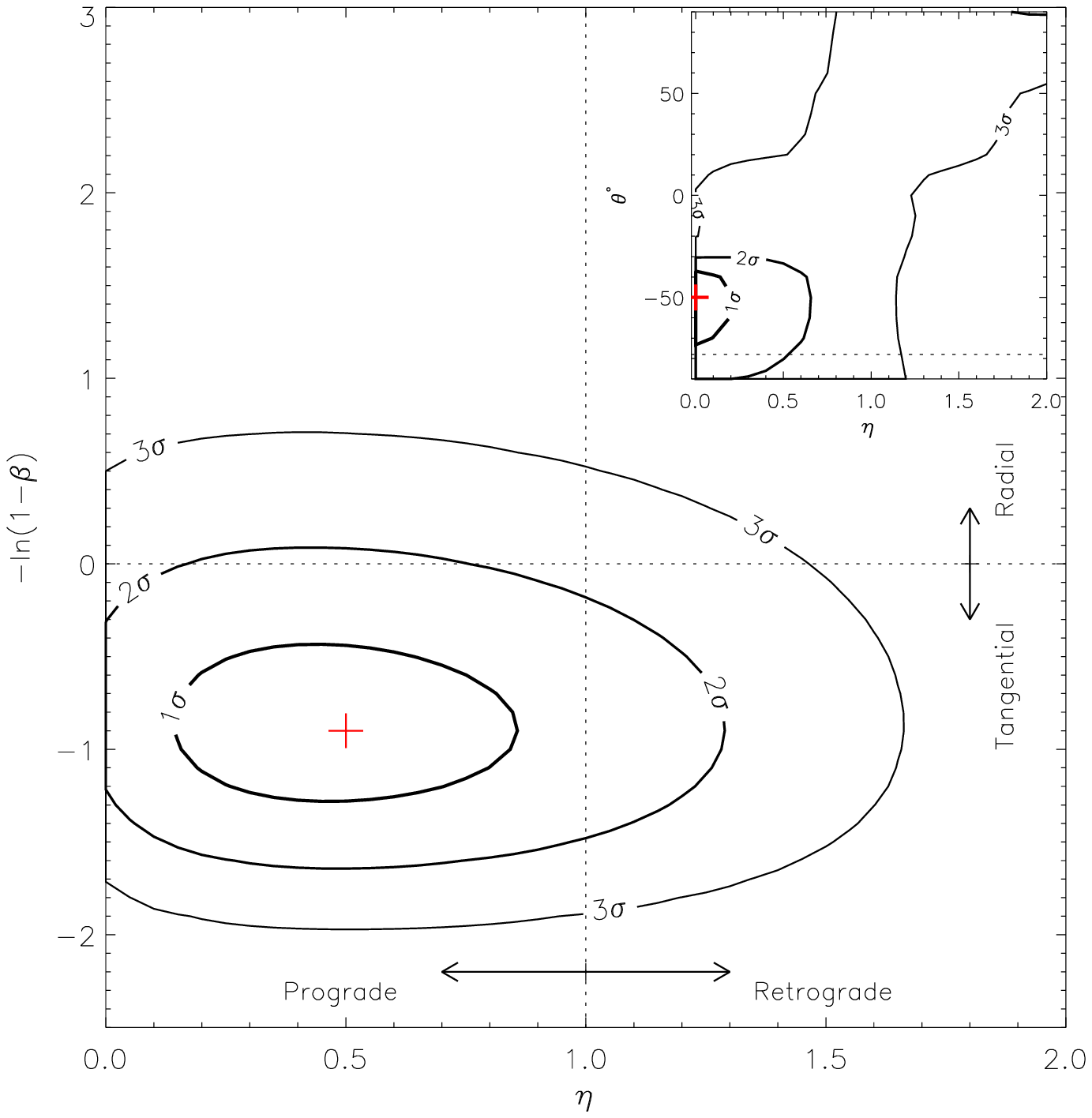}
  \end{minipage}
  \begin{minipage}{0.4 \linewidth}
    \centering
  \includegraphics[width=8cm, height=8cm]{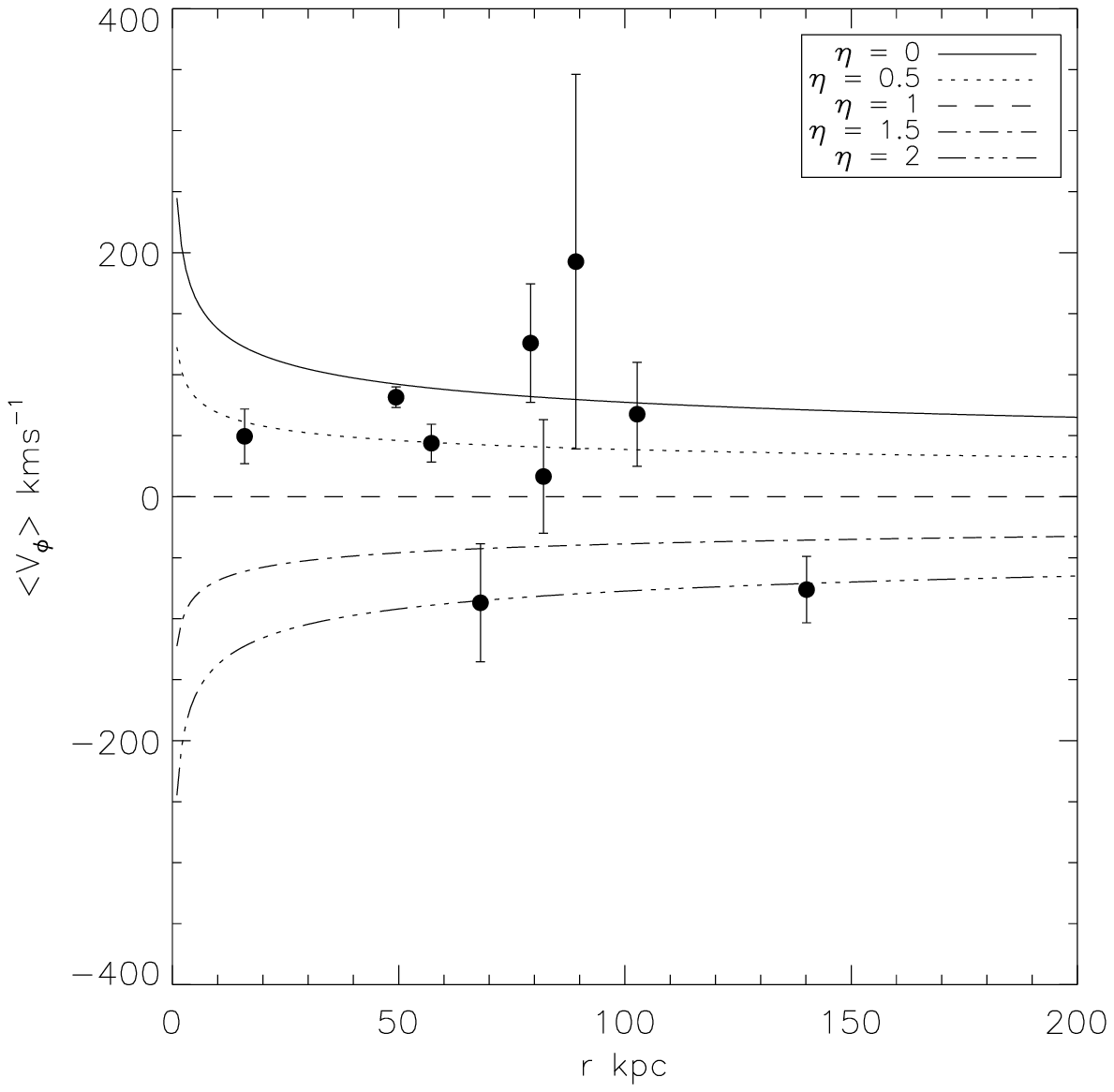}
    \end{minipage}
    \caption[MW satellites PM]{ \small Left: The 2D $\eta$, $\beta$
      confidence contour plane for the Milky Way satellite galaxies from the maximum likelihood
      analysis. Arrows illustrate
      the regions of prograde, retrograde, radial and tangential
      distributions respectively. The Milky Way satellites with
      available proper motions show evidence for prograde rotation
      ($\eta \sim 0.5$) and a tangential velocity distribution
      ($\beta \sim -1.5$). The inset of the figure shows the confidence
      contours when the normal vector of the angular momentum is
      tilted with respect to the Milky Way disc (by an angle
      $\theta$). More pronounced prograde rotation is seen when
      $\theta \approx -50^\circ$. The dotted line shows the
      approximate tilt angle found by \cite{metz07} using spatial
      coordinates of the satellites assuming they form a `disc of
      satellites'. Right: The streaming motion as a function of
      Galactocentric radius for a tangential distribution $\beta=-1.5$
      (see Fig.~\ref{fig:vphi}). Different linestyles represent
      different rotation parameters ($\eta$). Overplotted are the
      $v_\phi$ velocity components of the Milky Way satellites with
      available proper motions. Most exhibit positive $v_\phi$ values
      but Fornax and Ursa Minor have negative $v_\phi$.}
  \label{fig:mw_sat_pm}
\end{figure*}

To make further progress, we incorporate the available proper motion
data of the Milky Way satellites into the analysis. This restricts the
analysis to the classical satellites, as no proper motion measurements are
available for the ultra-faints. Fig.~\ref{fig:mw_sat_pm} (left panel) shows the
resulting likelihood distribution with respect to the rotation
parameter, $\eta$ and the anisotropy parameter, $\beta$.

With these extra constraints on the tangential velocity components,
the maximum likelihood method gives evidence for prograde rotation
($\eta=0.5^{+0.4}_{-0.2}$, $\beta=-1.5^{+0.7}_{-1.0}$). Evaluating
eqn~(\ref{eq:vphi}) gives $\langle v_\phi\rangle= (69 \pm 42) (r/10
\, \mathrm{kpc})^{-1/4} \, \mathrm{kms^{-1}}$. Averaging over the
volume interval ($10 < r/\mathrm{kpc} < 150$) gives $\langle
v_\phi\rangle= (38 \pm 23) \, \mathrm{kms^{-1}}$. The right hand panel of
Fig.~\ref{fig:mw_sat_pm} shows $v_\phi$ for the satellites with proper
motion data overplotted on a tangential model for $\langle v_\phi
(r)\rangle$ (see Fig.~\ref{fig:vphi}). Most of the satellites have
prograde rotation components, although the Fornax and Ursa Minor
satellites show retrograde rotation.

\cite{watkins10} recently reported $\beta =0.44$ for the Milky Way
satellites based on results of
simulations by \citealt{diemand07}. However, based on the proper
motion data, the same authors find $\beta \approx -4.5$, favouring
tangential orbits. Our result also favours tangential orbits (in
agreement with \cite{wilkinson99} who find $\beta \approx -1$).

The orbits of the satellites for which we have all velocity components
are preferentially polar ($v_{\theta}$ is the largest velocity
component). This has been noted in earlier work (e.g
\citealt{zaritsky99}) and has led to the suggestion that the Milky Way
satellites may occupy a highly inclined disc of satellites
(\citealt{palma02}; \citealt{kroupa05}; \citealt{metz07}). We have chosen the $L_z$ angular momentum
 component of the ensemble of satellites to be aligned with that of the
 disc of the Milky Way. As an inset in the left hand panel of Fig.~\ref{fig:mw_sat_pm},
we show how the apparent rotation changes if we introduce a tilted
angular momentum vector for the ensemble of satellite galaxies ($-90^\circ < \theta < 90^\circ$). Note that
$\theta=0$ corresponds to alignment  with the $L_z$ angular momentum
vector of the Milky Way disc. We find
more pronounced (prograde) rotation when $\theta =
-{50^\circ}^{+15}_{-20}$. This result is not surprising considering we
find preferentially polar orbits, but does not necessarily mean the
satellites occupy a rotationally supported disc. We also show the
approximate tilt angle found by \cite{metz07} of $\theta \approx
-78^\circ$ (dotted line)\footnote{There is a typo in the abstract of
  \cite{metz07} where an inclination angle of $88^\circ$ is given
  instead of the (correct) value of $78^\circ$} . The authors find the spatial distribution of
the satellites is best described by a highly inclined disc. Whilst our
results favour a slightly less inclined plane, it is interesting that
the deductions from both a kinematic and spatial analysis broadly
agree. However, without more accurate proper motion measurements the
disc of satellites hypothesis cannot be rigorously tested.

\section{The Local Standard of Rest}
\label{sec:discussion}

\begin{figure}
  \centering
  \includegraphics[width=8.5cm,height=8.5cm]{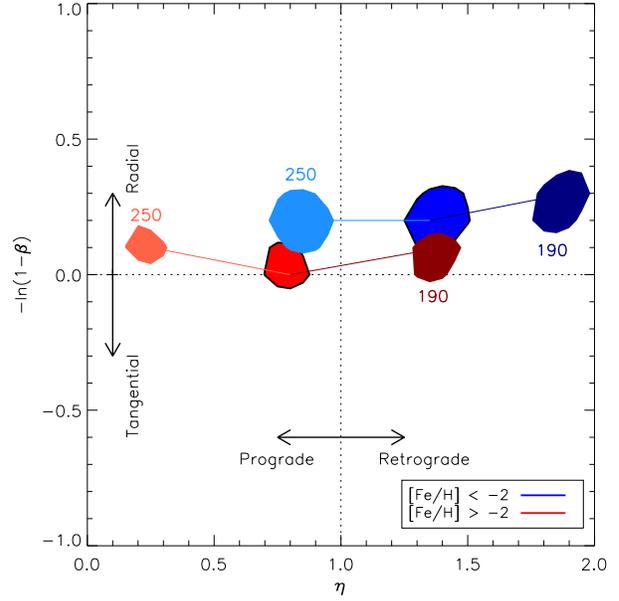}
  \caption[$L(\eta, \beta, r)$ MW BHBs]{\small As Fig. \ref{fig:bhb1}
    but we show the effects of varying the local standard of rest
    velocity ($\Theta_0$). Metal poor ($\mathrm{[Fe/H]} < -2$, $N=1135$) and
    metal `rich' ($ \mathrm{[Fe/H]} > -2$, $N=2125$) subsamples are shown by
    blue and red contours respectively. The paler shades show an
    \emph{increase} in the circular speed to $\Theta_0=250 \:
    \mathrm{kms^{-1}}$ whilst the darker shades show a
    \emph{decrease} in the circular speed to  $\Theta_0=190 \:\mathrm{kms^{-1}}$}
   
  \label{fig:bhb4}
\end{figure}

Our interpretation of any rotation signal in the Milky Way
stellar halo depends on our assumed local
standard of rest rotational velocity ($\Theta_0$). The IAU
(International Astronomical Union) recommend a value of
$\Theta_0=220 \: \mathrm{kms^{-1}}$. In practice, estimates of
the rotation speed in the literature vary between $184 \:
\mathrm{kms^{-1}}$ (\citealt{olling98}) and  $272 \:
\mathrm{kms^{-1}}$ (\citealt{mendez99}). Many of these estimates are
confined to regions within the solar neighbourhood and rely
on assuming a value for the Galactocentric distance
($R_0$). 

An abundance of evidence (net rotation, extent, metallicity, velocity, anisotropy,
dispersion profile) suggests the relatively metal-rich BHB population
of the Milky Way halo is associated with the accretion of a massive
satellite. The low metallicity and flat velocity dispersion profile of
the metal-poor population argue against an association with a single
accretion event. We infer that
this metal-poor population reflects the `primordial' metal-poor halo. The net
retrograde rotation may simply be the result of an \emph{underestimate} of
the local standard of rest rotational velocity. In Fig. \ref{fig:bhb4}
we show the effect of varying $\Theta_0$ on our results for the Milky
Way halo BHB population (see Fig. \ref{fig:bhb1}). Increasing our
adopted $\Theta_0$ value removes any retrograde signal in the
metal-poor population and enhances the net prograde signal for the
metal-rich population. We can provide functional forms for the
rotation signal of the two components by fitting a simple linear
relation to the adopted local standard of rest value:

\begin{equation}
\begin{split}
\langle v_\phi \rangle_{\mathrm{poor}}= \left(275 \,
\left[  \frac{\Theta_0}{220}\right]-300 \right) \: \mathrm{kms^{-1}} \\
\langle v_\phi \rangle_{\mathrm{rich}}=\left(330 \,
\left[\frac{\Theta_0}{220}\right]-315\right)\: \mathrm{kms^{-1}} \\
190 < \Theta_0 / \, \mathrm{kms^{-1}} < 250
\end{split}
\end{equation}

Here we have removed the dependence on Galactocentric radii of
$\langle v_{\phi} \rangle$ (see eqn \ref{eq:vphi}) by integrating over
the appropriate volume interval ($10 < r/\mathrm{kpc} < 50$). If we assume the metal-poor halo component has
no net rotation, then we can infer an independent estimate for the local
standard of rest rotational velocity of $\Theta_0 \approx 240 \: \mathrm{kms^{-1}}$.

An upward revision of the commonly adopted circular speed of $220 \:
\mathrm{kms^{-1}}$ agrees with recent work by \cite{reid09} and
\cite{bovy09} based on the kinematics of masers found in massive
star-forming regions of the Milky Way. \cite{reid09} fit for $R_0$ and
$\Theta_0$ simultaneously using kinematic and spatial information on
sources well beyond the local solar neighbourhood and established a
circular rotation speed, $\Theta_0 \approx 250 \:
\mathrm{kms^{-1}}$. \cite{bovy09} re-analysed the maser tracers using
less restrictive models and found a rotation speed of $\Theta_0
\approx 240-250 \: \mathrm{kms^{-1}}$. They confirm that there is no
conflict between recent determinations of the circular rotation speed
based on different methods (e.g. orbital fitting of the GD-1 stellar
stream, see \citealt{koposov10}) and quote a combined estimate of
$\Theta_0 \approx 240 \: \mathrm{kms^{-1}}$.

Note that \cite{carollo07} argued that the net retrograde signal in
the outer halo ($r > 10 \, \mathrm{kpc}$) is due to dynamical friction
effects. \cite{quinn86} show, based on the effects of dynamical
friction, that fragments on retrograde orbits surrender their orbital
energy to a much smaller extent than those on prograde orbits. The
disintegration of many of these fragments by tidal forces may result
in the formation of a halo stellar population with a net retrograde
asymmetry (e.g. \citealt{norris89}). Hence, in this picture the
metal-poor component is the accumulation of several (smaller)
accretion events. However, it is questionable whether this is a
plausible explanation seeing as dynamical friction mainly affects
massive satellites close to the plane of the disc. This is difficult
to reconcile with the aggregation of less massive satellites which
leave debris over a wide range of distances ($10 < r/\mathrm{kpc} <
50$) studied in this work.

We conclude that the lack of a plausible physical explanation for a
net retrograde rotating metal-poor halo population suggests that the
commonly adopted local standard of rest circular velocity needs to be
revised upwards. Assuming the primordial metal-poor halo has no net
rotation provides an independent estimate for this fundamental
Galactic parameter.

\section{Applications: The Andromeda galaxy}
We now apply our analysis to the Andromeda (M31) galaxy. In this case,
we transform to a coordinate system centred on M31. We define the
projected distances (in angular units) along the major and minor axis,
X and Y via:
\begin{equation}
\begin{split}
\mathrm{x_0= sin(\alpha-\alpha_0)cos(\delta)},\\
\mathrm{y_0= sin(\delta)cos(\delta_0)-cos(\alpha-\alpha_0)cos(\delta)sin(\delta_0)},\\
\mathrm{X= x_0 \, sin(\theta_a)+y_0 \, cos(\theta_a)},\\
\mathrm{Y= -x_0 \, cos(\theta_a)+y_0 \, sin(\theta_a)},
\end{split}
\end{equation}
where $\mathrm{x_0}, \mathrm{y_0}$ are Cartesian coordinates for
$\alpha$, $\delta$ with respect to the centre of M31
($\alpha_0=00^{\mathrm{h}}42^{\mathrm{m}}44.3^{\mathrm{s}} \:
\delta_0=+41^\circ16'09''$; \citealt{cotton99}) and
$\theta_a=37.7^\circ$ (\citealt{devac58}) is the position angle.  We
adopt the transformation outlined in Appendix A of \cite{evans94} to
relate the projected coordinates to the galaxy coordinates (assuming
an inclination angle, $i=77.5^\circ$).

\subsection{M31 Satellites}
\begin{table}
\begin{center}
\renewcommand{\tabcolsep}{0.1cm}
\renewcommand{\arraystretch}{0.2}
  \begin{tabular}{  l  l  r  r  r  r  c}
    \hline 
     Name & Type & $l  \,(^\circ)$ & $b \, (^\circ)$ & $d$ (kpc) &
    $v_{\rm h}$ kms$^{-1}$ & Ref\\
    \hline
    M31 & Spiral & 121.2 & -21.6 & $785^{+25}_{-25}$ & -301 &  1 
    \\
    \\
    And I* & dSph & 121.7 & -24.8 & $745^{+24}_{-24}$ & -380  & 1 
    \\
    \\
    And II* & dSph & 128.9 & -29.2 & $652^{+18}_{-18}$ & -188  & 1 
    \\
    \\
    And III* & dSph & 119.4 & -26.3 & $749^{+24}_{-24}$ & -355  & 1 
    \\
    \\
    And V & dSph & 126.2 & -15.1 & $774^{+28}_{-28}$ & -403  & 1 
    \\
    \\
    And VI* & dSph & 106.0 & -36.3 & $783^{+25}_{-25}$ & -354  & 1 
    \\
    \\
    And VII & dSph & 109.5 & -9.9 & $763^{+25}_{-25}$ & -307  & 1 
    \\
    \\
    And IX* & dSph & 123.2 & -19.7 & $765^{+5}_{-150}$ & -207.7  & 2 
    \\
    \\
    And X* & dSph & 125.8 & -18.0 & $702^{+36}_{-36}$ & -163.8  & 3,4 
    \\
    \\
    And XI* & dSph & 121.7 & -29.0 & $760^{+10}_{-150}$ & -419.6  & 2 
    \\
    \\
    And XII* & dSph & 122.0 & -29.5 & $830^{+170}_{-30}$ & -558.4  & 2 
    \\
    \\
    And XIII* & dSph & 123.0 & -29.9 & $910^{+30}_{-160}$ & -195.0  & 2 
    \\
    \\
    And XIV* & dSph & 123.0 & -33.2 & $740^{+110}_{-110}$ & -481.1 & 5 
    \\
    \\
    And XV* & dSph & 127.9 & -24.5 & $770^{+70}_{-70}$ & -339 & 6 
    \\
    \\
    And XVI & dSph & 124.9 & -30.5 & $525^{+50}_{-50}$ & -385 & 6 
    \\
    \\
    Pisces & dIrr/dSph & 126.8 & -40.9 & $769^{+23}_{-23}$ & -286 & 1 
    \\
    \\
    Pegasus & dIrr/dSph & 94.8 & -43.6 & $919^{+30}_{-30}$ & -182 & 1 
    \\
    \\
    NGC 147 & dE & 119.8 & -14.3 & $675^{+27}_{-27}$ & -193 & 1 
    \\
    \\
    IC 1613 & dE & 129.8 & -60.6 & $700^{+35}_{-35}$ & -232 & 1 
    \\
    \\
    NGC 185 & dE & 120.79 & -14.5 & $616^{+26}_{-26}$ & -202 & 1 
    \\
    \\
    NGC 205 & dE & 120.7 & -21.1 & $824^{+27}_{-27}$ & -244 & 1 
    \\
    \\
    IC 10 & dIrr & 119.0 & -3.3 & $825^{+50}_{-50}$ & -344 & 1 
    \\
    \\
    M32 & cE & 121.2 & -22.0 & $785^{+25}_{-25}$ & -205 & 1 
    \\
    \hline
  \end{tabular}
  \caption[M31 Satellites]
  {Properties of the Andromeda Satellites: Here, we give the satellite
    name and type (dSph = dwarf spheroidal, dIrr = dwarf irregular, dE
    = dwarf elliptical and cE = classical elliptical), Galactic
    coordinates ($l,b$), heliocentric distance $d$ and heliocentric
    velocity $v_{\rm h}$. Satellites belonging to the rotating group
    are starred. References: (1) \citealt{mcconnachie06},
    (2) \citealt{collins09}, (3) \citealt{zucker07}, (4)
    \citealt{kalirai09},(5) \citealt{majewski07}, (6) \citealt{letarte09}}
\label{tab:and_sat}
\end{center}
\end{table}

\begin{figure*}
  \begin{minipage}{0.55 \linewidth}
    \centering
  \includegraphics[width=9cm,height=9cm]{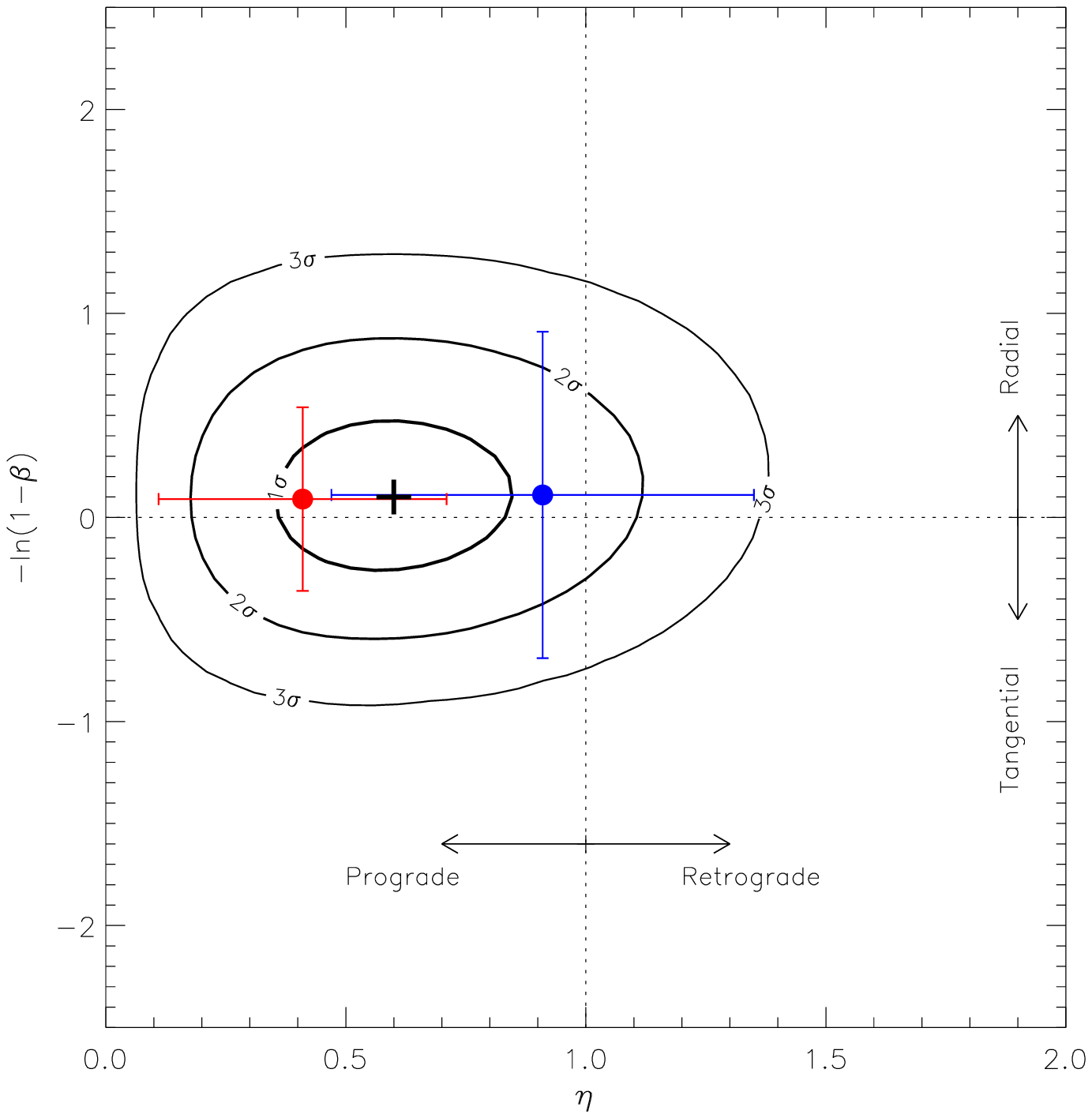}
  \end{minipage}
  \begin{minipage}{0.4 \linewidth}
    \centering
  \includegraphics[width=8cm,height=8cm]{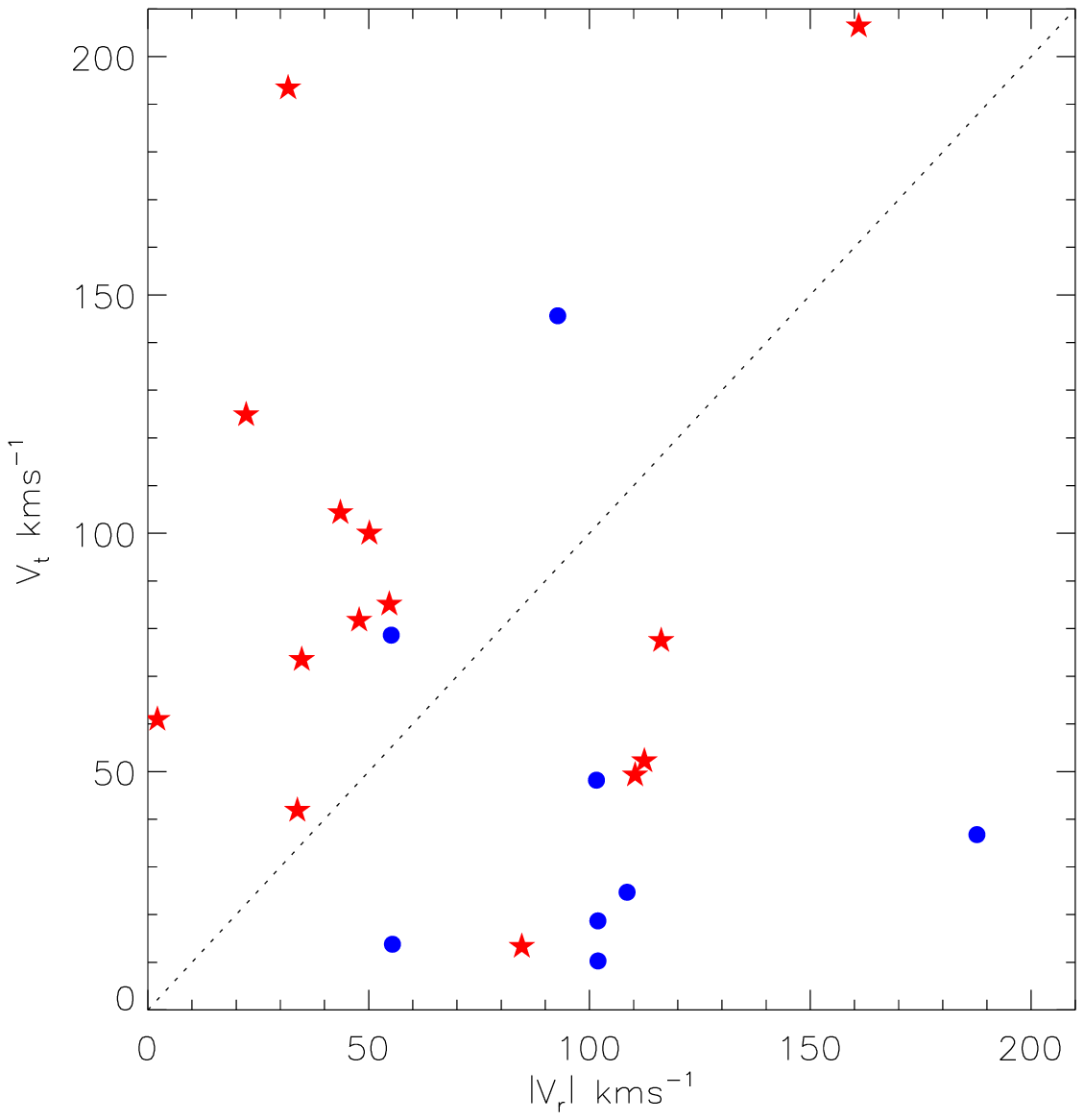}
  \end{minipage}
  \caption[M31 satellites]{Left: The confidence contours of the
  orbital parameters $\eta$ and $\beta$ for the M31 satellites. The
  red and blue points correspond to the dSphs (rotating) and non-dSph (non-rotating)
  subsets respectively. Arrows represent the regions of prograde,
  retrograde, radial and tangential distributions. Right: The line of
  sight velocities of the M31 satellites decomposed into tangential
  and radial components. These are lower limits as only one component
  of velocity is known. The red stars and blue circles represent the
  members of the dSph and non-dSph groups respectively.}
    \label{fig:and_all}
\end{figure*}

The main drawback in the M31 analysis is that the distances to the
satellites are not as well constrained as for the Milky Way satellites. To take
this into account, we integrate the distribution function along the
line of sight between the distance errors.  The left hand panel of
Fig.~\ref{fig:and_all} shows the results of the maximum likelihood
analysis applied to all of the M31 satellites (see Table
~\ref{tab:and_sat}). Unlike the situation for the Milky Way
satellites, our line of sight to M31 is not constrained to a
predominantly radial velocity component. Fig.~\ref{fig:and_all} shows
evidence of prograde rotation from the line of sight velocity
alone. We further investigate this rotating component by splitting the
M31 sample into two groups based on the satellite type. These are
shown by the red (dSph) and blue points (non-dSph) in Fig.~\ref{fig:and_all} where we find a
rotating group (dSph, red) and a non-rotating group (non-dSph, blue). The rotating
subset has a rotation parameter, $\eta=0.4^{+0.3}_{-0.3}$ and
anisotropy parameter, $\beta=0.1^{+0.5}_{-0.5}$, which corresponds to
a streaming motion, $\langle v_\phi\rangle= (62 \pm 34) (r/10 \,
\mathrm{kpc})^{-1/4} \, \mathrm{kms^{-1}}$. Averaging over a suitable
volume interval ($10 < r/\mathrm{kpc} < 200$) gives  $\langle v_\phi\rangle= (32 \pm 17) \, \mathrm{kms^{-1}}$.

The observed line of sight velocities of the M31 satellites can be
decomposed into radial and tangential components by making use of the
spatial information relative to M31. These are lower limits as we only
have the line of sight velocity component (see
\citealt{metz07}). There are also large uncertainties on the distances
of the satellites which permeate into the spatial coordinates with
respect to M31. However, we still find it instructive to look at these
approximate velocity components. The right panel of
Fig.~\ref{fig:and_all} shows the lower limit tangential and radial
velocity components for the M31 satellites. The members of the dSph rotating group are shown by red
stars and the non-rotating satellites are shown by the blue
circles. The majority of the rotating subgroup
have a larger tangential velocity than radial velocity component. 

The rotating subgroup consists entirely of dwarf spheroidal galaxies. Satellites with similar accretion histories can share
common properties such as their `type' (i.e dSph) and orbital
orientation. The apparent rotation of the dSph satellites of M31
suggests that these dwarfs may share a common origin.

\subsection{M31 globular clusters}

Over recent years, studies by several authors
(e.g. \citealt{perrett02}, \citealt{galleti07}, \citealt{kim07},
\citealt{caldwell09}) have produced a rich sample of M31 globular
clusters with available kinematic data. Our data derives from the
Revised Bologna Catalogue (\citealt{rbc}). Distance measurements are
unavailable for the globular clusters, so we integrate our
distribution function along the line of sight.

\begin{figure}
  \centering
  \includegraphics[width=8.5cm, height=8.5cm]{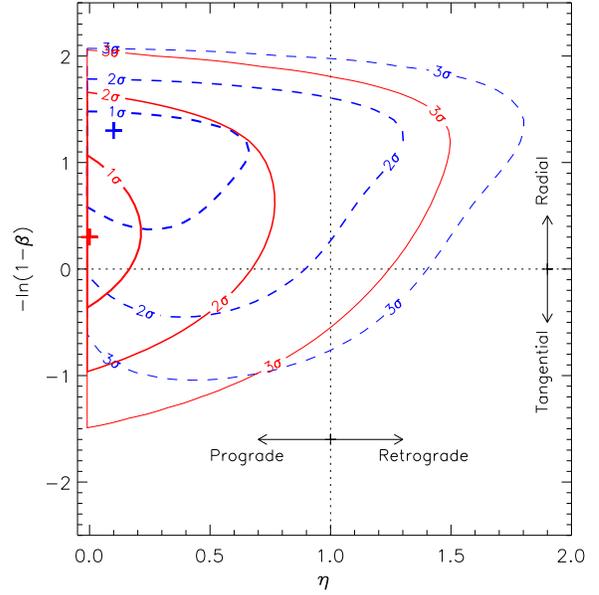}
  \caption[$L(/eta, /beta)$ M31 GCs]{\small The confidence contours
    of the orbital parameters $\eta$ and $\beta$ for the M31 globular
    clusters. We only consider globular clusters where $|Y| > 5 \: \mathrm{kpc}$ to
    probe the outer regions and reduce disc contamination. The solid
    red and dashed blue contours represent the metal-rich
    ([Fe/H] $>$ -1) and metal-poor  ([Fe/H] $<$ -1) subsamples respectively}  
  \label{fig:and_gc}
\end{figure}

We consider the 93 globular clusters for which the projected distance
along the semi-minor axis is greater than 5 kpc ($|Y| > 5 \:
\mathrm{kpc}$). This criterion was introduced by \cite{lee08}.  For
54 of these clusters, metallicities are available and so we further
divide the sample into 24 metal-rich ($[\mathrm{Fe/H}] > -1$) and 30
metal-poor ($[\mathrm{Fe/H}] < -1$) components.  Fig.~\ref{fig:and_gc}
shows the outer M31 globular clusters are rotating prograde with
respect to the M31 disc. The metal-rich subset (solid red contours)
has $\eta=0^{+0.2}_{-0.2}$ and $\beta=0.3^{+0.4}_{-0.6}$ giving a streaming
motion, $\langle v_\phi\rangle= (94 \pm 35) (r/10 \,
\mathrm{kpc})^{-1/4} \,
\mathrm{kms^{-1}}$. The metal-poor subset (dashed blue contours) has a
less pronounced rotation with $\eta=0.1 \pm 0.5$, and a more
pronounced radial anisotropy, $\beta=0.7^{+0.1}_{-0.2}$. This indicates
that it has a smaller streaming velocity: $\langle v_\phi\rangle= (53 \pm 34)
(r/10 \, \mathrm{kpc})^{-1/4} \, \mathrm{kms^{-1}}$. However, the broad
overlap of the 1$\sigma$ error envelopes of these samples mean that it
is premature to conclude that we are indeed seeing a significant
difference between the metal-rich and metal-poor populations. \cite{lee08} also find
strong rotation for the outer globular clusters, suggesting a hot
rotating halo or an extended bulge system in M31. The role of the halo
in M31 has been contentious. Some studies have suggested M31 has an
extended bulge but no halo (e.g. \citealp{pritchet94, hurley-keller04,
  irwin05, merrett06}), whilst more recent studies have revealed
evidence for a huge halo (\citealp{chapman06, kalirai06,
  ibata07}). The weaker rotational component in the metal-poor sample
may suggest a transition from the bulge to a halo component but a
larger kinematic sample extending further along the minor-axis is
required for confirmation. 

%% file: conclusions.tex
\section{Conclusions}

We studied the rotational properties of halo populations in the Milky
Way Galaxy and Andromeda (M31). We modelled the gravitational
potential, $\Phi$, and the density profile of the tracer population,
$\rho$, as spherically symmetric power-laws with indices $\gamma$ and
$\alpha$ respectively. The shape of the velocity distribution is
controlled by two parameters, namely $\beta$ which governs the
velocity anisotropy and $\eta$ which governs the rotation.

For our applications, we used the potential $\Phi \sim r^{-\gamma}$
with $\gamma = 0.5$ as this is a good approximation to a
Navarro-Frenk-White profile at large radii. Our tracer populations
have a density profile $\rho \sim r^{-\alpha}$ with typically
$\alpha=3.5$.  A maximum likelihood method is used to constrain the
orbital parameters $\beta$ and $\eta$ for five different populations:
(1) Milky way halo blue horizontal branch (BHB) stars, (2) Milky Way
globular clusters, (3) Milky Way satellites, (4) M31 satellites and
(5) M31 globular clusters. We summarise our conclusions as follows:

\bigskip
\noindent
(1) From our constructed sample of halo BHB stars from SDSS, we find
evidence for two different halo populations: a prograde, rotating,
comparatively metal-rich component ($[\mathrm{Fe/H}] > -2$) and a
retrograde, rotating, comparatively metal-poor component
($[\mathrm{Fe/H}] < -2$). The mean streaming motions are $\langle
v_\phi\rangle= (21 \pm 11) (r/10 \, \mathrm{kpc})^{-1/4}\,
\mathrm{kms^{-1}}$ and $\langle v_\phi\rangle= -(35 \pm 10) \, (r/10 \,
\mathrm{kpc})^{-1/4}\, \mathrm{kms^{-1}}$ for the richer and poorer
populations respectively. A rough
estimate suggests that these structures may have stellar masses of $M
\sim 10^{8}M_{\odot}$. We find evidence from the velocity dispersion
profile that the metal-rich sample is associated with the accretion of
a massive satellite. The line of sight velocity dispersion of the
metal-rich sample has a kinematically cold component at $20 <
r/\mathrm{kpc} < 30$, which could be evidence for a shell like
structure caused by the accretion event. We find no such evidence to
suggest the metal-poor sample is associated with an accretion event. We
suggest that this population typifies the primordial stellar halo and
the net retrograde rotation reflects an underestimate in our adopted
local standard of rest circular velocity, which may be as high as
$\sim 240 \: \mathrm{kms^{-1}}$.
 
\bigskip
\noindent
(2) The halo globular clusters of the Milky Way show evidence of net
prograde rotation, $\langle v_\phi\rangle= (58 \pm 27)
(r/10\,\mathrm{kpc})^{-1/4}\, \mathrm{kms^{-1}}$, and have a mildly
radial velocity distribution. Previous studies have found a mild net
prograde rotation from an analysis of the line of sight velocities of
the halo globular clusters (e.g. \citealt{frenk80}). Our results are
in agreement, but this deduction arises from the analysis of those 15 of the
41 halo globular clusters which have available proper motions. We note that the majority of our
globular cluster sample have metallicities in the range $-2 <
\mathrm{[Fe/H]} < -1$ and are located between $10 < r/\mathrm{kpc} <
20$. It is interesting that we also see net prograde rotation in the
halo BHB stars in this distance and metallicity range. This suggests
the relatively metal-rich BHBs and halo globular clusters may share a
similar formation history.

\bigskip
\noindent
(3) Line of sight velocities alone poorly constrain the orbital
parameters of the Milky Way satellites. Using available proper motions
for 9 of the classical satellites, we find evidence for prograde
rotation, $\langle v_\phi\rangle= (69 \pm 42) (r/10\,
\mathrm{kpc})^{-1/4}\, \mathrm{kms^{-1}}$, and a tangential anisotropy
parameter $\beta=-1.5^{+0.7}_{-1.0}$. The orbits are preferentially
polar. This leads to the detection of a more pronounced rotation when
the normal vector of rotation is inclined by $|\theta| \approx 50^\circ$
to the normal of the disc. This hints that the satellites may be part
of a rotationally supported disc (as postulated by a number of
authors), but without more accurate proper motions this hypothesis
cannot be confirmed.

\bigskip
\noindent
(4) Application of these methods to the M31 satellites is potentially powerful, as the line of sight velocity has contributions from
both the tangential and radial velocity components with respect to the
M31 centre. The overall population shows hints of prograde
rotation. We can divide the satellites into two groups. One possesses
pronounced prograde rotation, $\langle v_\phi\rangle= (62 \pm 34)
(r/10\, \mathrm{kpc})^{-1/4}\, \mathrm{kms^{-1}}$, whilst the other
shows no evidence for a rotating signal. Interestingly, the rotating
group consists entirely of the dwarf spheroidal satellites. The correlation between satellite type and orbital
orientation suggests these satellites may share a common origin.

\bigskip
\noindent
(5) There is a rich dataset on the M31 globular clusters which remains
relatively unexploited. We examined the globular clusters with
projected distance along the minor-axis greater than 5 kpc. These
outer globular clusters have prograde rotation which is more
pronounced for the metal-rich ([Fe/H] $>$ -1) subset. We find $\langle
v_\phi\rangle= (94 \pm 35) (r/10\, \mathrm{kpc})^{-1/4}
\mathrm{kms^{-1}}$ for the metal-rich population and $\langle
v_\phi\rangle= (53 \pm 34) (r/10\, \mathrm{kpc})^{-1/4}
\mathrm{kms^{-1}}$ for the metal-poor population. The strong rotation,
especially in the metal-rich sample, suggests these globular clusters
may belong to the M31 bulge system. However, a more robust explanation awaits a more substantial
kinematic sample extending further along the minor-axis.

\bigskip
\noindent
The \emph{Gaia} satellite promises to provide proper motion data on
all dwarf galaxies of the Milky Way and M31 and for thousands of halo
stars. \cite{wilkinson99} suggested that \emph{Gaia} can constrain the
proper motion of distant classical dwarfs in the Milky way, like Leo I,
to within $\pm 15 \: \mathrm{kms^{-1}}$ and nearer dwarfs like Ursa
Minor to within $\pm 1 \: \mathrm{kms^{-1}}$. Whilst \emph{Gaia} will
improve the proper motion measurements for the classical satellites,
deeper complements to \emph{Gaia} such as the LSST will be needed to
achieve this level of accuracy for the ultra-faint satellites.

\begin{figure}
  \centering
  \includegraphics[width=8cm, height=8cm]{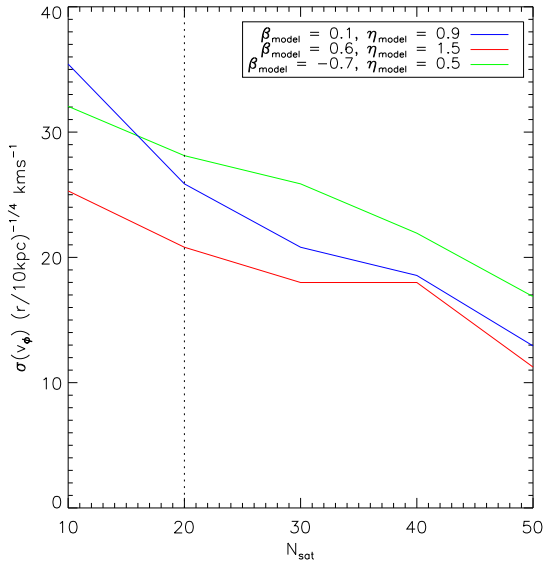}
  \caption[$\sigma_{v_{\phi}}$, $N_{\mathrm{sat}}$]{\small The
    estimated error in the mean streaming motion of the Milky Way
    satellites as a function of number of satellites. We assume random
    errors on the tangential velocity components in the range $1 \,
    \mathrm{kms^{-1}} \le \sigma \le 15 \, \mathrm{kms^{-1}}$. Three different
      models are used, shown by the blue ($\eta=0.9$, $\beta=0.1$), red
      ($\eta=1.9$, $\beta=0.6$) and green ($\eta=0.5$, $\beta = -0.7$) lines respectively.}
  \label{fig:gaia}
\end{figure}

In anticipation of this future work, we close by
considering Monte Carlo simulations of our distribution functions to
estimate how well we can constrain rotation with these future
astrometric surveys. We assume random errors on the tangential
velocity components between $1 \, \mathrm{kms^{-1}}$ and $15 \:
\mathrm{kms^{-1}}$. Fig.~\ref{fig:gaia} gives a rough estimate of how
the accuracy depends on the
number of satellites with proper motion measurements. For $\mathrm{N}
\approx 20$ satellites, we can constrain the mean streaming motion to
$\sigma_{v_{\phi}} \approx 30 (r/10 \, \mathrm{kpc})^{-1/4} \,
\mathrm{kms^{-1}}$. This evaluates to $\sigma_{v_{\phi}} \approx (10 -
20) \, \mathrm{kms^{-1}}$ for satellites between $r= (50 - 250) \,
\mathrm{kpc}$.
The combination of more accurate proper motion measurements with
radial velocities can potentially probe any apparent kinematic
dichotomy between the classical and ultra-faint satellites, and
perhaps provide more robust evidence bearing on the disc of
satellites.

Our analysis of the Milky Way stellar halo using BHB stars will be
greatly improved by increasing the number of targets at larger
radii. Coupled with more accurate radial velocity measurements, we
will be able to map the kinematic structure of the stellar halo in
greater detail and see if the apparent dual nature of the metal-rich
and metal-poor components continues out to further reaches of the
halo. LSST will hopefully play a vital role in finding these BHB
targets.

